\documentclass[12pt]{article}
\usepackage{amsfonts,amssymb,amsmath,mathtools,graphics,slashed}
\usepackage{hyperref,hep,graphicx,subfig}
\usepackage{rotating}
\usepackage{xcolor}
\usepackage{verbatim}
\usepackage[final]{showlabels}

\definecolor{greenish}{RGB}{0,190,0}

\definecolor{yellowish}{RGB}{190,190,0}

\definecolor{bluish}{RGB}{0,0,190}

\newcommand{\nn}{\notag \\}

\makeatletter

\@addtoreset{equation}{section}
\makeatother

\begin{document}

\begin{titlepage}

\vfill


\vfill

\begin{center}
   \baselineskip=16pt
   {\Large\bf Holographic Dissipation from the Symplectic Current}
  \vskip 1.5cm
  \vskip 1.5cm
      Aristomenis Donos$^1$, Polydoros Kailidis$^1$ and Christiana Pantelidou$^2$\\
   \vskip .6cm
      \begin{small}
      \textit{$^1$ Centre for Particle Theory and Department of Mathematical Sciences,\\ Durham University,
       Durham, DH1 3LE, U.K.}\\
        \textit{$^2$ School of Mathematics and Statistics,\\
University College Dublin, Belfield, Dublin 4, Ireland}
        \end{small}\\   
         
\end{center}

\vfill

\begin{center}
\textbf{Abstract}
\end{center}
\begin{quote}
We develop analytic techniques to construct the leading dissipative terms in a derivative expansion of holographic fluids. Our basic ingredient is the Crnkovic-Witten symplectic current of classical gravity which we use to extract the dissipative transport coefficients of holographic fluids, assuming knowledge of the thermodynamics and the near horizon geometries of the bulk black hole geometries. We apply our techniques to non-conformal neutral fluids to reproduce previous results on the shear viscosity and generalise a known expression for the bulk viscosity.
\end{quote}

\vfill

\end{titlepage}

\setcounter{equation}{0}

\section{Introduction}

One of the most important problems in theoretical physics is to understand the collective behaviour of large systems at finite temperature and chemical potential. This task is particularly difficult for systems whose constituents are strongly interacting and lacking a weakly interacting quasiparticle description \cite{Hartnoll:2016apf}.

At large scales, such systems are expected to be governed by the conserved quantities of global symmetries and the Goldstone modes associated to their breaking \cite{Kovtun:2012rj}. One of the main goals of such an effective description is to flesh out the broad properties of long wavelength dynamics and package all the microscopic details of interactions in a set of transport coefficients which are functions of the thermal state.

The holographic duality provides a framework where non-trivial questions about strongly interacting non-gravitational quantum systems can be mapped to significantly simpler problems in theories which contain gravity \cite{Aharony:1999ti,Witten:1998qj}. In a certain large $N$ and strong coupling limit, these theories admit a semiclassical description in which the gravitational side becomes classical. In this regime, the thermal states from the field theory side get mapped onto black hole solutions with the temperature set by the event horizon. This powerful duality has allowed the study of phases of strongly coupled matter with symmetries which are broken either spontaneously \cite{Hartnoll:2008vx,Gubser:2008px,Gubser:2008zu,Nakamura:2009tf,Donos:2011bh} or explicitly \cite{Horowitz:2012ky,Donos:2013eha,Andrade:2013gsa,Donos:2014yya,Donos:2021pkk,Ammon:2021pyz}.

In the classical limit of holographic theories, one can perform  real time microscopic computations around the thermal states and in which a hydrodynamic limit is not necessary. In this framework, it is therefore possible to study the validity and eventually the breakdown of hydrodynamics \cite{Heller:2013fn,Heller:2015dha,Withers:2018srf,Grozdanov:2019kge,Grozdanov:2019uhi}. The most well studied case is the breakdown of hydrodynamics as a function of the wavelength of fluctuations. In this scenario, one of the hydrodynamic poles collides with one of the poles of a UV degree of freedom which is gapped. Holography provides an invaluable laboratory where the non-hydrodynamic modes of the UV theory can be retained in the description.

Another approach where holography becomes particularly useful concerns the effective theory governing the hydrodynamics modes itself. By considering its hydrodynamic limit, holography can be used to either derive or confirm the effective theory itself from first principles \cite{Son:2007vk,Bhattacharyya:2008jc,Haack:2008cp,Erdmenger:2008rm,Banerjee:2008th}. In this approach, one expects that the corresponding transport coefficients should be expressible in terms of quantities related to the black hole geometries dual to the thermal states \cite{Iqbal:2008by,Donos:2014cya,Donos:2015gia}.

Significant progress towards this direction  has been made over the previous years. The first quantity that was computed holographically was the shear viscosity of hologrpahic theories \cite{Son:2007vk} by direct computation of the retarded Green's around the AdS-Schwarzschild black brane solution which is known analytically. A significant improvement was later achieved in \cite{Iqbal:2008by} where the shear viscosity and the electric conductivity of holographic theories at zero chemical potential was expressed in terms of black hole horizon data without relying on the knowledge of the analytic form of the gravitational solution. This was achieved by exploiting the existence of a conserved electric current in the bulk theory allowing one to relate boundary to horizon quantities. This approach is natural for the case of the electric conductivity since the bulk gauge field which is dual to the field theory electric current satisfies Maxwell's equations. However, the authors of \cite{Son:2007vk} had already indicated that the shear viscosity could be extracted in a similar way after performing a dimensional reduction and reducing the problem to solving Maxwell's equations in a lower dimensional spacetime.

In more recent years, in the context of applications of holography to condensed matter systems, it became natural to study the thermoelectric properties of holographic matter at finite temperature and chemical potential. For the thermoelectric conductivities to be finite in the zero frequency limit, momentum should be non-conserved so that it can relax.  This implies that translational symmetry has to be explicitly broken, leaving only time translations as a symmetry for the bulk geometries. A general recipe for computing the matrix of the DC thermoelectric conductivities was given in a series of papers \cite{Donos:2014cya,Donos:2015gia,Donos:2015bxe,Donos:2017oym} after exploiting the existence of time translations as a symmetry in the bulk geometry and relating thermoelectric currents that one can define on the black hole horizon to the field theory currents on the conformal boundary.

A common feature of the progress reported above exploits the Kubo formulae to extract the relevant transport coefficients. Here, we will develop analytic techniques that will allow us to write directly a set of constitutive relations for the stress tensor of neutral relativistic fluids in terms of a local temperature and fluid velocity. Our methods are general enough that we don't have to rely on the explicit knowledge of the background black hole spacetimes apart from some of their general properties. As part of our derivation, we will reproduce the famous formula for the shear viscosity of \cite{Son:2007vk,Iqbal:2008by}.

Moreover, by introducing additional scales to the problem via relevant scalar operators, our fluids will also have a non-trivial bulk viscosity. For a given class of theories, a formula for the bulk viscosity had been computed in the past \cite{Eling:2011ms} purely from the point view of the black hole horizon fluid \cite{Eling:2009pb}. In our work, we will read off the bulk viscosity from the stress tensor of the boundary theory after performing a change of hydrodynamic variables that will bring us to the so called Landau frame.

One of the key ingredients of our method is the Crnkovic-Witten symplectic current \cite{Crnkovic:1986ex} which can be thought of as a generalisation of Liouville's theorem to classical gravitational and gauge theories. In the past, it has found many interesting applications which include the minisuperspace quantisation of large families of supersymmetric solutions of higher dimensional supergravities which are known as bubbling solutions \cite{Grant:2005qc,Maoz:2005nk,Donos:2005vs}. Another, slightly different application is the thermodynamics and the first law of black holes \cite{Wald:1993nt,Iyer:1994ys,Papadimitriou:2005ii}.

Here, we will use the Crnkovic-Witten symplectic form in order to derive the leading dissipative corrections to ideal hydrodynamics for a holographic relativistic fluid. The important feature of the symplectic current is that it is antisymmetric in the space of perturbative solutions and that it is divergence free. Antisymmetry makes the symplectic current expandable for perturbations which are infinitesimally close to each other. This is certainly the case when one of the solutions used to construct the symplectic current is a static thermodynamic perturbation and the second one is a hydrodynamic perturbation whose infinite wavelength limit coincides with the former one. The fact that the symplectic current is divergence free allows us to use Stokes' theorem and relate boundary quantities to the black hole horizon. The purpose of this paper is to explain in detail how the procedure works in such a simple example. Note that this is independent of the existence of conservation laws and symmetries which have been exploited in the past in the general context of the membrane paradigm as Liouville's theorem is always true.

The paper is organised in five main sections. In section \ref{sec:setup} we introduce the class of our holographic models along with the relevant thermodynamics and the important Ward identities that will later be part of our hydrodynamic description. In section \ref{sec:sympl_current} we introduce the symplectic current along with its relevance to holography in the hydrodynamic limit. We also describe the main elements of strategy in order to extract the stress tensor of the boundary theory. In section \ref{sec:pert} we discuss the relevant bulk perturbations for our system. Moreover, we discuss the thermodynamic and boost perturbations which will be the starting point of our derivative expansion appearing towards the end of the section. In section \ref{sec:const_relations} we derive the constitutive relations for our holographic fluid and we read off the shear and bulk viscosities.  In section \ref{sec:numerics} we carry out numerical checks for our bulk viscosity formula and we conclude with a discussion in section \ref{sec:discussion}.

\newpage

\section{Holographic Setup}\label{sec:setup}

In this section we will present the setup in which we wish to study holographic relativistic fluids at finite temperature. In the first subsection we will discuss the relevant gravitational models along with the black holes which are dual to the thermal states we wish to perturb. In the second subsection we will review the important ingredients of holographic renormalisation and the thermodynamic properties of our background geometries.

\subsection{The Holographic Model}

Here we will introduce the class of holographic theories we will use to model our non-conformal fluids. Apart from the bulk metric, we will include a number $n_S$ of bulk scalar fields $\varphi_I$, dual to the field theory operators $\mathcal{O}_{I}$. Without loss of generality, we will assume that the bulk theory is described by the  action,
\begin{align}\label{eq:bulk_action}
S_b=\int\,d^{4}x\,\sqrt{-g}\,\left(R-\frac{1}{2}\Phi_{IJ}\,\nabla_\mu\varphi^I\, \nabla^\mu\varphi^J-V\right)\,,
\end{align}
where the functions $\Phi^{IJ}$ and the potential $V$ can, in general, depend on the scalar fields $\varphi_I$. The above action gives rise to the equations of motion,
\begin{align}\label{eq:eoms}
R_{\mu\nu}-\frac{1}{2}g_{\mu\nu}\,V-\frac{1}{2}\Phi_{IJ}\,\nabla_\mu\varphi^I\,\nabla_\nu\varphi^J=&0\,,\notag\\
\nabla_\mu\left(\Phi_{KJ}\,\nabla^\mu\varphi^J \right)-\frac{1}{2}\partial_{\varphi_K}\Phi_{IJ}\,\nabla_\mu\varphi^I\, \nabla^\mu\varphi^J-\partial_{\varphi_K}V=&0\,.
\end{align}

In order for the geometry to asymptote to $AdS_4$ of unit radius, we will assume that for small values of the scalar fields the functions that appear in our bulk action \eqref{eq:bulk_action} behave as,
\begin{align}
V&\approx -6+\frac{1}{2}\sum_I m_I^2\,(\varphi^I)^2+\cdots\,,\notag\\
\Phi_{IJ}&\approx \delta_{IJ}+\cdots\,.
\end{align}
As usual, the bulk mass parameters $m_I^2$ determine the conformal dimensions $\Delta_I$ of the dual operators $\mathcal{O}_I$ according to $\Delta_I\,(\Delta_I-3)=m_I^2$. In order to break conformality, it will be important for us that $n_R<n_S$ of our scalar operators are relevant with respect to the UV theory having conformal dimensions $1/2<\Delta_I<3$, $I=1,\cdots,n_R$. In this case, we can introduce constant deformation parameters $\varphi_{(s)}^I$ which preserve the Lorentz symmetry of the dual theory.

The black brane backgrounds capturing the thermal states we will be interested in are captured by the ansatz,
\begin{align}\label{eq:background}
ds^{2}&=-U(r)\,dt^{2}+\frac{dr^{2}}{U(r)}+e^{2g(r)}\,\left(dx^{2}+dy^{2} \right)\,,\nn
\phi^I&=\phi^I (r)\,.
\end{align}
Near the conformal boundary, our functions will admit the expansions,
\begin{align}\label{eq:uv_bcs}
U(r)&\approx (r+R)^2+\cdots+\frac{g_{(v)}}{r+R}+\cdots,\qquad g(r)\approx \ln(r+R)+\cdots\,,\notag\\
\varphi^I(r)&\approx \frac{\varphi_{(s)}^I}{(r+R)^{3-\Delta_I}}+\cdots +\frac{\varphi_{(v)}^I}{(r+R)^{\Delta_I}}+\cdots\,.
\end{align}
The deformation parameters $\varphi_{(s)}^I$ are part of our fixed conditions on the conformal boundary, while $\varphi_{(v)}^I$ fix the VEVs $\langle\mathcal{O}_I\rangle$ of the dual field theory operators. The constant of integration $R$ in \eqref{eq:uv_bcs} is chosen so that the horizon of our black brane geometry is fixed at $r=0$. Close to the horizon, regularity imposes the analytic expansion,
\begin{align}\label{eq:nh_bcs}
U(r)&\approx 4\pi T r+\cdots,\qquad g(r)=g_{(0)}+g_{(1)}\,r+\cdots,\notag\\
\varphi^I&\approx \varphi^I_{(0)}+\cdots\,,
\end{align}
where $T$ is the Hawking temperature of the event horizon and is kept fixed. The constants of integration $g_{(0)}$, $\varphi^I_{(0)}$, $g_{(v)}$ and $\varphi^I_{(v)}$ are fixed by solving the equations of motion \eqref{eq:eoms}, subject to the above boundary conditions.

It is important to note that some of the scalar operators might take a VEV spontaneously as we lower the temperature of the system. This is allowed to happen since our UV theory is assumed to have been deformed by at least a relevant operator, with the deformation parameter setting a scale and therefore breaking scaling invariance at low energies.

\subsection{Thermodynamics and Ward Identities}
An important aspect of our paper is that we will be able to extract quantities which are directly relevant to the field theory living on the conformal boundary. In order to do this, the bulk action \eqref{eq:bulk_action} needs to be supplemented by suitable counter terms that render it finite \cite{Skenderis:2002wp}. Equally important is the fact that these counter terms will make the variational problem well defined \cite{Papadimitriou:2005ii}. For simplicity, here we list a few universal terms which are necessary,
\begin{align}\label{eq:bdy_action}
S_{bdr}=&-\int_{\partial M}d^{3}x\,\sqrt{-\gamma}\,\left(-2K + 4 +R_{bdr}\right)\notag\\
&\quad -\frac{1}{2}\int_{\partial M}d^{3}x\,\sqrt{-\gamma}\,\sum_I[(3-\Delta_I)(\varphi^{I})^2-\frac{1}{2\Delta_{I}-5}\,\partial_{a}\varphi^I\,\partial^{a}\varphi^I]+\cdots\,,
\end{align}
where $\gamma_{ab}$ is the induced metric on the hypersurface $\partial M$ of constant radial coordinate $r$. The Ricci and extrinsic curvature scalars of $\gamma_{ab}$ are $R_{bdr}$ and $K$ correspondingly.

To obtain the free energy of our black hole spacetimes \eqref{eq:background}, we need to compute the total action $I_{Tot}=I_b+I_{bdr}$ for the Euclidean backgrounds with Wick rotated time $t=-i\,\tau$. The free energy density $w_{FE}$ will then be given by $w_{FE}=T\,I_{Tot}$. After exploiting the existence of the Killing vector $\partial_t$ for the background solutions \eqref{eq:background}, the bulk part of the action can be written as a total derivative,
\begin{align}\label{eq:action_bulk_contr}
I_b=\frac{1}{T}\,\int_{0}^\infty\,dr\,\left(e^{2g}\,U^\prime \right)^\prime\,.
\end{align}
The free energy density and the energy density $\epsilon$ of our system are related by the usual Legendre transformation,
\begin{align}
w_{FE}=\epsilon-T\,s\,,
\end{align}
where $s$ is the entropy density arising from the horizon contribution of the integral \eqref{eq:action_bulk_contr}. We also note here that our black holes satisfy the enlarged first low of thermodynamics,
\begin{align}
dw_{FE}=-s\,dT-\langle\mathcal{O}_I\rangle\,d\varphi_{(s)}^I\,.
\end{align}

In terms of the horizon data appearing in the expansion \eqref{eq:nh_bcs} we have,
\begin{align}
s=4\pi\,e^{2\,g_{(0)}}\,.
\end{align}
This is the Bekenstein-Hawking formula for the entropy relating the entropy $s$ to the volume density of the horizon. Combining the asymptotic expansion \eqref{eq:uv_bcs} with \eqref{eq:action_bulk_contr} and \eqref{eq:bdy_action} we obtain the energy,
\begin{align}
\epsilon=-2\,g_{(v)}-\varphi_{(s)}^I\,\langle \mathcal{O}_I\rangle\,,
\end{align}
in terms of the constants of integration on the boundary with the scalar VEV $\langle \mathcal{O}_I\rangle=(2\Delta_I-3)\,\varphi_{(v)}^I$.

By varying the total action $S=S_b+S_{bdr}$ with respect to the metric on the conformal boundary $\gamma_{ab}$, we can obtain the expectation value of the field theory stress tensor $T_{ab}$.  At this point, it is useful to discuss a version of the bulk action which only contains first order derivatives. After integrating by parts the Einstein-Hilbert terms in \eqref{eq:bulk_action}, and dropping the horizon contribution, the term at infinity cancels the Gibbons-Hawking term in the boundary terms of equation \eqref{eq:bdy_action}. This yields a first order action which takes the form\footnote{Note that in order to obtain this form of the action, we have also dropped a term coming from the horizon of the black holes. For the static backgrounds, this term gives the entropy term in the expression for the grand canonical free energy. However, for calculations in real time this term is dropped.},
\begin{align}\label{eq:first_order_action}
S=\int\,d^4x\, \mathcal{L}\left(g_{\mu\nu},\partial_\lambda g_{\mu\nu},\varphi^I,\partial_\lambda\varphi^I\right)+\tilde{S}_{bdr}\,.
\end{align}
As we explain in Appendix \ref{app:fo_action}, the new boundary action now only contains the counterterms which are necessary to regularise the bulk action and which also define the correct variational problem. More concretely, the universal terms read,
\begin{align}\label{eq:bdyp_action}
\tilde{S}_{bdr}=&-\int_{\partial M}d^{3}x\,\sqrt{-\gamma}\,\left(4 +R_{bdr}\right)\notag\\
&\quad -\frac{1}{2}\int_{\partial M}d^{3}x\,\sqrt{-\gamma}\,\sum_I[(3-\Delta_I)(\varphi^{I})^2-\frac{1}{2\Delta_{I}-5}\,\partial_{a}\varphi^I\,\partial^{a}\varphi^I]+\cdots\,.
\end{align}
The above allows us to write the expression for the VEVs of the boundary stress tensor and the scalar operators,
\begin{align}
\langle T^{\mu\nu}\rangle&=\lim_{r\to\infty}\frac{2\,r^5}{\sqrt{-\gamma}}\left[ \frac{\partial\mathcal{L}}{\partial(\partial_r g_{\mu\nu})}+\frac{\delta \tilde{S}_{bdr}}{\delta\gamma_{\mu\nu}}\right]\,,\nn
\langle \mathcal{O}_I\rangle&=\lim_{r\to\infty} \frac{r^{\Delta_I}}{\sqrt{-\gamma}}\left[\frac{\partial\mathcal{L}}{\partial(\partial_r \varphi^I)}+ \frac{\delta \tilde{S}_{bdr}}{\delta \varphi^I}\right]\,,
\end{align}
which will become useful when we discuss our use of the gravitational symplectic current in context of holography.

The gravitational constraints which one can obtain by foliating the bulk spacetime by constant $r$ hypersufaces can be imposed anywhere as the equations of motion guarantee that they will be satisfied everywhere. Choosing the hypersurface close to the conformal boundary yields the Ward identities,
\begin{align}\label{eq:Ward}
\nabla_a \langle T^{ab}\rangle=\nabla^{b}\varphi_{(s)}^I\,\langle \mathcal{O}_I\rangle\,.
\end{align}

In our context, the boundary metric and the background sources preserve translations. Moreover, as we will later see, the pertrurbative sources we will include are going to be order $\mathcal{O}(\varepsilon)$ in the derivative expansion. In order to extract the leading order corrections to the ideal fluid, we will need to know the boundary stress tensor only up to order $\mathcal{O}(\varepsilon^2)$. This simple observation allows us to drop the derivative terms that come from varying the boundary action \eqref{eq:bdyp_action} leading to,
\begin{align}
\langle T^{\mu\nu}\rangle&\approx\lim_{r\to\infty}\frac{2\,r^5}{\sqrt{-\gamma}}\left[ \frac{\partial\mathcal{L}}{\partial(\partial_r g_{\mu\nu})}-\sqrt{-\gamma}\left(2-\frac{1}{4}\,(\Delta_I-3)\,(\varphi^I)^{2} \right)\gamma^{\mu\nu} \right]+\mathcal{O}(\partial^2)\,,\nn
\langle \mathcal{O}_I\rangle&\approx\lim_{r\to\infty} \frac{r^{\Delta_I}}{\sqrt{-\gamma}}\left[\frac{\partial\mathcal{L}}{\partial(\partial_r \varphi^I)}+\sqrt{-\gamma}\,(\Delta_I-3)\,\varphi^I\right]+\mathcal{O}(\partial^2)\,.
\end{align}
From the above, we see that for our purposes it is only the algebraic counterterms that will contibute to hologrpahic renormalisation at the order of the derivative expansion we will be working.

It is useful to note that the non-trivial expectation values of the stress tensor components are given by,
\begin{align}\label{eq:bac_stress}
\langle T_{tt}\rangle=\epsilon\,,\quad \langle T_{xx} \rangle= \langle T_{yy} \rangle=p\,,
\end{align}
where $p$ is the pressure of our thermal state. Given that $\partial_x$ and $\partial_y$ are Killing vectors of our geometry, we also have the relation $p=-w_{FE}$ between the pressure and the free energy. This result can be shown by either exploiting the Komar integrals or one can give an argument similar to that of \cite{Donos:2013cka} for such geometries.

\section{The Symplectic Current Strategy}\label{sec:sympl_current}
In this section we will introduce the Crnkovic-Witten symplectic form \cite{Crnkovic:1986ex} focusing on our application to the hydrodynamic expansion. It has already been used in similar applications the past \cite{Donos:2021pkk,Donos:2022xfd} and here we will highlight the important ingredients and strategy for the present work. To define it, we consider a generic classical Lagrangian field theory of a collection $\varphi^I$ of fields and two perturbative solutions $\delta_1\phi^I$ and $\delta_2\phi^I$ around a background $\phi^I_b$. If the Lagrangian density $\mathcal{L}(\phi^I,\partial_\mu\phi^I)$ can be written in terms of the fields and their first derivatives then the vector density,
\begin{align}\label{eq:scurrent_def}
P^\mu_{\delta_1,\delta_2}=\delta_1\phi^I\,\delta_2\left(\frac{\partial\mathcal{L}}{\partial \partial_\mu\phi^I} \right)-\delta_2\phi^I\,\delta_1\left(\frac{\partial\mathcal{L}}{\partial \partial_\mu\phi^I} \right)\,,
\end{align}
is divergence free,
\begin{align}\label{eq:div_free}
\partial_\mu P^\mu_{\delta_1,\delta_2}=0\,.
\end{align}
Given the bulk action \eqref{eq:bulk_action}, the contributing terms to the derivatives of the bulk action in \eqref{eq:scurrent_def} are given by,
\begin{align}\label{eq:sympl_cur_contr}
\frac{\partial\mathcal{L}}{\partial\partial_\mu g_{\alpha\beta}}&=\sqrt{-g}\,\Gamma^\mu_{\gamma\delta}\left(\,g^{\gamma\alpha}\,g^{\delta\beta}-\frac{1}{2}\,g^{\gamma\delta}g^{\alpha\beta} \right)-\sqrt{-g}\,\Gamma^\kappa_{\kappa\lambda}\,\left(g^{\mu\left(\alpha\right.}g^{\left.\beta\right)\lambda}-\frac{1}{2}\,g^{\mu \lambda}g^{\alpha\beta}\right)\,,\notag\\
\frac{\partial\mathcal{L}}{\partial\partial_\mu\varphi^I}&=-\sqrt{-g}\,\Phi_{IJ}\,\partial^\mu\varphi^J\,.
\end{align}

The symplectic current \eqref{eq:scurrent_def} is antisymmetric in field space and as such, when the second perturbative solution is infinitesimally close to the first one,  the symplectic form is expandable around zero. In our work, the role of this parameter will be played by the wavenumber of the hydrodynamic perturbations which we will take to be parametrically small, of order $\mathcal{O}(\varepsilon)$.

As we will see in the next sections, the role of one of the two solutions needed to construct the symplectic current \eqref{eq:scurrent_def} will be played by a static solution. The second solution will always be the hydrodynamic perturbations we wish to study and which are infinitesimally close to thermodynamic perturbations. As we will argue, the latter can be obtained from the background black holes of equation \eqref{eq:background} by either varyring with respect to temperature by $\delta T\sim \mathcal{O}(\varepsilon)$ or by performing  coordinate transformations corresponding to a Lorentzian boost from the boundary theory perspective. In the limit we will be working, the boost parameters will be given by a fluid velocity parameter $\delta v^i\sim\mathcal{O}(\varepsilon)$.

In general, both the static (known) solutions as well as the hydrodynamic perturbations we wish to construct will contain perturbative sources. The sources of the hydrodynamic perturbations will be suppressed in the $\varepsilon$ expansion. Our main strategy will be to exploit the fact that the radial component of the symplectic \eqref{eq:scurrent_def} will asymptote to,
\begin{align}\label{eq:scurrent_asymptotics}
P^r_{\delta_1,\delta_2}=&\frac{1}{r^3}\,\left(\delta_1\varphi_{(s)}^I\,\delta_2\left(\sqrt{-\gamma}\,\langle\mathcal{O}_I\rangle\right)-\delta_2\varphi_{(s)}^I\,\delta_1\left(\sqrt{-\gamma}\,\langle\mathcal{O}_I\rangle\right)\right)\nn
&+\frac{1}{r^3}\frac{1}{2}\left(\delta_1\gamma_{ab}\,\delta_2\left(\sqrt{-\gamma}\,\langle T^{ab}\rangle\right)-\delta_2\gamma_{ab}\,\delta_1\left(\sqrt{-\gamma}\,\langle T^{ab}\rangle\right)\right)+\cdots
\end{align}
for all the pairs of solutions we will consider, at leading order in the hydrodynamic expansion. In the non-hydrodynamic limit, the above expression will have to be supplemented by the non-trivial contribution of the derivative terms in \eqref{eq:bdyp_action}. However, these terms would not contain any additional information in terms of the VEVs and would be fixed entirely by the sources.

The above shows that by integrating the divergence free condition  \eqref{eq:div_free}, we can relate the asymptotic form \eqref{eq:scurrent_asymptotics} to horizon quantities and bulk integrals. Therefore, by choosing appropriate static seed solutions, we can draw conclusions about the VEVs of operators along the hydrodynamic perturbations we wish to study. As we will see in detail, the bulk integrals will eventually drop out in the hydrodynamic limit, leading to expressions relating the VEVs of the stress tensor components to horizon quantities.

Finally, the last step will be to obtain closed expressions for the stress tensor of our theory in terms of $\delta T$ and $\delta v^i$.  By choosing source free perturbations to construct the symplectic current \eqref{eq:scurrent_def}, we will be able to express all the near horizon constants of integration as functions of temperature fluctuations $\delta T$ and the local fluid velocity $\delta v^i$. This step will give us the constitutive relations for our hydrodynamics after fixing an appropriate fluid frame. These relations will turn the Ward identy \eqref{eq:Ward} to a closed system of equations that will allow us to determine the linear response of our system against external sources.

\section{Perturbations}\label{sec:pert}

In this section we will construct the hydrodynamic perturbations of our relativistic system. Before considering their hydrodynamic limit, we will set up the more general framework of perturbations  which can depend on the field theory coordinates $\left(t,x^i\right)$. The fluctuations we are interested in will be around the static backgrounds of equation \eqref{eq:background} which also enjoy translational invariance.

This allows for a Fourier decomposition of perturbations according to,
\begin{align}\label{eq:fourier_modes}
\delta\mathcal{F}(t,x^i;r)=e^{-iw\,(t+S(r))+ik_i x^i}\,\delta f(r)\,,
\end{align}
where $\delta\mathcal{F}$  represents perturbations of the scalars as well as the metric field components. Moreover, by choosing the function $S(r)$ to approach $S(r)\to\frac{\ln r}{4\pi T}+\cdots$ close to the black hole horizon at $r=0$, we are guaranteed to the correct infalling boundary conditions provided that $\delta f(r)$ admits a Taylor series expansion there. Finally, in order for the holographic dictionary to be solely dictated by the asymptotics of $\delta f(r)$, we will choose $S(r)$ to behave as $S(r)\to \mathcal{O}(1/r^3)$ close to the conformal boundary.

Our perturbations will then be governed by the asymptotic expansions,
\begin{align}\label{eq:pert_uv_bcs}
\delta g_{ab}(r)&=(r+R)^2\,\left(\delta s_{ab}+\cdots +\frac{\delta t_{ab}}{(r+R)^3}+\cdots\right)\,,\nn
\delta g_{ra}(r)&=\mathcal{O}\left(\frac{1}{r^3}\right),\quad\delta g_{rr}(r)=\mathcal{O}\left(\frac{1}{r^4} \right)\,,\nn
\delta\varphi^I(r)&=\frac{\delta\varphi_{(v)}^I}{(r+R)^{\Delta_I}}+\cdots\,,
\end{align}
where the constants of integration $\delta s_{ab}$ correspond to perturbative sources for the stress tensor of the boundary theory. The constants of integration $\delta t_{ab}$ and $\delta\varphi_{(v)}^I$ will fix the VEVs of the stress tensor and scalar operators correspondingly. Note that we prefer to not fix a particular coordinate system everywhere in the bulk. We will only fix the asymptotic behaviour of our coordinate system through the expansion \eqref{eq:pert_uv_bcs}.

On the other side of the geometry, near the horizon at $r=0$, we want to impose infalling boundary conditions which can be achieved through the expansions,
\begin{align}\label{eq:gen_exp}
\delta g_{tt}(r)&= 4\pi T\,r\, \delta g_{tt}^{(0)}+\cdots\,,\quad
\delta g_{rr}(r)=\frac{\delta g_{rr}^{(0)}}{4\pi T\,r}+\cdots\, \,,\nn
\delta g_{ti}(r)&=\delta g_{ti}^{(0)}+r\,\delta g_{ti}^{(1)}+\cdots\,,\quad 
\delta g_{ri}(r)=\frac{\delta g_{ri}^{(0)}}{4\pi T\,r}+\delta g_{ri}^{(1)}+\cdots\,,\nn
\delta g_{ij}(r)&=\delta g_{ij}^{(0)}+\cdots\,,\quad\quad
\delta g_{tr}(r)=\delta g_{tr}^{(0)}+\cdots\,,\nn
\delta \varphi^{I}(r)&=\delta \varphi^{I(0)}+\cdots \,.
\end{align}
In order to achieve regularity, the above need to be supplemented by
\begin{align}\label{eq:nh_reg}
-2\pi T(\delta g_{tt}^{(0)}+\delta g_{rr}^{(0)})=-4\pi T\,\delta g_{rt}^{(0)}&\equiv \delta T_h\,,\notag\\
\delta g_{ti}^{(0)}=\delta g_{ri}^{(0)}&\equiv-\delta u_{i}\,.
\end{align}
where we have defined a sense of a local temperature $\delta T_h$ and fluid velocity $\delta u_{i}$ on the black hole horizon. These will not be directly relevant to us since we will discuss the fluid from the boundary point of view.

In the next subsections, we will consider the hydrodynamic perturbations of the system. We will think of these as deformations of the static thermodynamic perturbations we will consider in subsection \ref{sec:pert_therm}. In subsection \ref{sec:pert_diff} we will consider all possible static solutions which can be obtained by large coordinate transformations in the bulk and which we will use in order to introduce hydrodynamic sources in our system.

\subsection{Thermodynamic Perturbations}\label{sec:pert_therm}

In this subsection we will discuss the source free static solutions which will be the infinite wavelength limit of our hydrodynamic perturbations. The first example is the solution we can obtain by varying the temperature of the background black holes \eqref{eq:background} according to,
\begin{align}\label{eq:static_pert_dT}
\delta_T g_{tt}&=-\partial_T U\,,\quad \delta_T g_{rr}=-\frac{\partial_T U}{U^2}\,,\quad \delta_T g_{tr}=U\,\partial_T S^\prime\,,\nn
\delta_T g_{ra}&=\delta_T g_{tx}=\delta_T g_{ty}=\delta_T g_{xy}=0\,,\quad \delta_T g_{xx}=\delta_T g_{yy}=2\,e^{2g}\,\partial_T g\,,\nn
\delta_T\varphi&=\partial_T\varphi\,.
\end{align}
Notice that apart from a simple partial derivative with respect to temperature, we have also performed an infinitesimal coordinate transformation,
\begin{align}
t\to t-\partial_T S\,\delta T\,,
\end{align}
in order to satisfy the infalling boundary conditions of equations \eqref{eq:gen_exp} and \eqref{eq:nh_reg} with $\delta p=4\pi\,\delta T$ and $v_i=0$. From the boundary point of view, the perturbation generated by varying the temperature will yield the non-trivial stress tensor components,
\begin{align}
\delta_T\langle T^{tt}\rangle&=\partial_T\epsilon=T\,\partial_T s\,,\nn
\delta_T\langle T^{ij}\rangle&=\delta^{ij}\,\partial_T p=\delta^{ij}\,s\,.
\end{align}

The second static solution we would like to consider is generated by boundary infinitesimal Lorentz boosts with parameter $\delta v_i$. From the boundary point view, this corresponds to the isomorphism,
\begin{align}
t\to t- \delta v_i\,x^i\,,\quad x^i\to x^i-\delta v^i\,t\,.
\end{align}
In order to obtain a regular solution in the bulk, we need to see the above as the asymptotics of a large coordinate transformation which is otherwise regular everywhere in the bulk. This is achieved by the coordinate transformation,
\begin{align}
t\to t- \delta v_i\,x^i\,,\quad x^i\to x^i-\delta v^i\,(t+S(r))\,,
\end{align}
leading to the bulk perturbation,
\begin{align} \label{eq:static_pert_dv}
\delta_{v_j} g_{tt}&=\delta_{v_j} g_{rr}=\delta_{v_j} g_{tr}=0\,,\quad \delta_{v_j}\varphi=0\,,\nn
\delta_{v_j} g_{ti}&=\delta^j_i\,\left(U-e^{2g}\right)\,,\quad \delta_{v_j} g_{ri}=-\delta^j_i\,e^{2g}\,S^\prime\,.
\end{align}
The above perturbation for the metric satisfies the infalling boundary conditions \eqref{eq:gen_exp} and \eqref{eq:nh_reg} with $\delta T_h=0$ and $\delta u_i=e^{2g^{(0)}}\,\delta v_i$. The above leads to the following perturbation for the components of the boundary stress tensor,
\begin{align}
\delta_{v_j}\langle T^{ti}\rangle =\delta_{v_j}\langle T^{it}\rangle=(\epsilon+p)\,\delta^{ij}\,.
\end{align}

\subsection{Large Differomorhisms}\label{sec:pert_diff}

In the previous subsections we discussed source free static perturbations which will naturally lead to black hole quasinormal modes when promoted to hydrodynamic perturbations. Here, we will discuss static perturbations which will contain sources for the stress tensor from the field theory point of view.

A natural way to achieve this is to consider large coordinate transformations which are everywhere regular and which alter the form of the asymptotic metric. This is in contrast with the Lorentz boosts of the previous section which asymptote to Killing vector of the boundary metric.

From the boundary point of view, it is easy to see that we can obtain coordinate independent metric deformations through the simple coordinate transformations,
\begin{align}\label{eq:coord_trans_sources}
x^a\to x^a+ \delta s^a{}_b\,(x^b+\delta^b_t\,S(r))\,,
\end{align}
leading to the boundary metric perturbations,
\begin{align}
\delta g_{ab}=\eta_{ac}\,\delta s^c{}_b+\eta_{bc}\,\delta s^c{}_a=2\,\delta s_{\left(ab\right)}\,.
\end{align}
The extra term in the bracket of \eqref{eq:coord_trans_sources}, is there to guarantee regularity close to the horizon at $r=0$. It will later be useful to list  the precise form of the perturbations we obtain from the inequivalent coordinate transformations,
\begin{align}\label{eq:static_pert_source}
\delta_{s_{tt}}g_{tt}&=2\,U\,,\quad \delta_{s_{tt}}g_{tr}=U\,S^\prime\,,\nn
\delta_{s_{tx}}g_{tx}&=U\,,\nn
\delta_{s_{xt}}g_{tx}&=e^{2g}\,,\quad \delta_{s_{xt}}g_{rx}=e^{2g}\,S^\prime\,,\nn
\delta_{s_{xx}}g_{xx}&=2\,e^{2g}\,,\nn
\delta_{s_{xy}}g_{xy}&=e^{2g}\,.
\end{align}
In the above we have omitted the obvious transformations corresponding to $\delta s_{ty}$ and $\delta s_{yy}$ as well as the components of the perturbations which remain trivial.

From the previous discussion, there seem to be two ways to introduce a static source $\delta\gamma_{ti}$ for the boundary metric, depending on whether $\delta s_{ti}$ or $\delta s_{it}$ is non-zero. However, the two are connected by a source-free Lorentz boost.  In order to introduce appropriate spacetime dependent sources for our boundary theory later, we will only use the transformation which has $\delta s_{ti}=0$. The reason is that this transformation preserves the Cauchy surfaces of constant time from the field theory point of view. Its only effect is to change the vector of the time flow with respect these Cauchy surfaces. In other words, at zero frequency and wavevector, the source $\delta \gamma_{ti}$ will reduce to a Galilean boost. Another way to see this is to realise the perturbation generated by $\delta s_{ti}$ is nothing but a linear combination of the perturbation generated by $\delta s_{it}$ with that generated by the boost $\delta v_i$.

We will close this subsection by summarising the perturbations for the VEVs of the stress tensor components,
\begin{align}
\delta_{s_{tt}}\langle T^{tt}\rangle&=2\,\epsilon\,,\nn
\delta_{s_{tx}}\langle T^{tx}\rangle&=p\,,\nn
\delta_{s_{xt}}\langle T^{tx}\rangle&=-\epsilon\,,\nn
\delta_{s_{xx}}\langle T^{xx}\rangle&=-2\,p\,,\nn
\delta_{s_{xy}}\langle T^{xy}\rangle&=-p\,,
\end{align}
where $p$ is the pressure of the thermal state.

\subsection{Hydrodynamic Perturbations}\label{sec:pert_hydro}

In the previous subsections we considered static perturbations which are independent of the field theory coordinates. In this section we will use those as seed solutions in order to construct perturbations which depend weakly on the boundary coordinates. In order to introduce this time dependence, we will use the Fourier mode decomposition \eqref{eq:fourier_modes} with $k_i=\varepsilon\,q_i$ and $\varepsilon$ a small number which will serve as our hydro expansion parameter. Similarly, we will extract a factor of $\varepsilon$ from the frequency according to $w=\varepsilon\,\omega$ since at exactly $\varepsilon=0$ we expect to recover the static modes we discussed in subsections \ref{sec:pert_therm} and \ref{sec:pert_diff}.

This observation also allows us to write an ansatz for the $\varepsilon$ bulk hydrodynamic modes according to,
\begin{align}\label{eq:hydro_pert_exp}
\delta_H f&=\varepsilon\,\delta f^{(1)}+\varepsilon^2\,\delta f^{(2)}+\cdots\nn
&=\delta_T f\,\delta T+\delta_{v_i} f\,\delta v_i+\delta_{s_{ab}} f\,\delta s_{ab}+\varepsilon^2\,\delta f^{(2)}+\cdots\,,
\end{align}
where we took the temperature variation $\delta T$, the fluid velocity parameter $\delta v^i$ and the sources $\delta s_{ab}$ to all be of order $\varepsilon$. In the language of hydrodynamics, the leading piece $\delta f^{(1)}$ constructed from the static solutions, will play the role of the ideal part of the fluid. The corrections $\delta f^{(2)}$ are the important bit we are missing in order to supplement the ideal hydrodynamics we have written in the previous sections with the first dissipative corrections. This is one of the aspects where the symplectic current of section \ref{sec:sympl_current} will be useful.

At this point we can write the constitutive relations for the stress tensor fluctuations,
\begin{align}\label{eq:const_rels}
\delta\langle T^{tt}\rangle&=T\,\partial_T s\,\delta T+2\,\epsilon\,\delta s^{tt}+\varepsilon^2\,\delta\langle T^{tt}\rangle_{(2)}+\cdots\,,\nn
\delta\langle T^{it}\rangle&=\delta\langle T^{ti}\rangle=(\epsilon+p)\,\delta v^i+\epsilon\,\delta s^{it}+\varepsilon^2\,\delta\langle T^{ti}\rangle_{(2)}+\cdots\,,\nn
\delta\langle T^{ij}\rangle&=\delta^{ij}\,s\,\delta T-2p\,\delta s^{\left(ij\right)}+\varepsilon^2\,\delta\langle T^{ij}\rangle_{(2)}+\cdots\,,
\end{align}
where $\delta\langle T^{ab}\rangle_{(2)}$ are the dissipative pieces of the stress tensor which are not captured by the ideal part.

In particular, using the hydrodynamic perturbation \eqref{eq:hydro_pert_exp} as one of the two perturbative solutions for the symplectic current along with the solutions of section \ref{sec:pert_diff} and \ref{sec:pert_hydro}, we will be able to write the leading dissipative corrections to the stress tensor in terms of the fluctuations $\delta T$ and $\delta v_i$. After fixing a specific fluid frame, we will see that the corresponding transport coefficients can be expressed in terms of the black hole horizon data entering in the near horizon expansion \eqref{eq:nh_bcs}.

It is worth noting that the ideal part of the constitutive relations in equations \eqref{eq:const_rels} is universal.  When used in the Ward identities \eqref{eq:Ward}, they yield the leading part of the fluid equations of motion,
\begin{align}\label{eq:eoms_ideal}
\partial_t \delta T&=-\frac{s}{\partial_T s}\,\delta^{ij}\left(\partial_i(\delta v_j -\delta s_{jt})+\partial_t\delta s_{ij}\right)+\mathcal{O}(\varepsilon^3)\,,\nn
\partial_t \delta v_i&=\partial_i\delta s_{tt}-T^{-1}\,\partial_i\delta T+\mathcal{O}(\varepsilon^3)\,.
\end{align}
With the sources set to zero, the above equations yield the two sounds modes and the shear mode which are familiar from relativistic hydrodynamics. The dispersion relations at leading order are,
\begin{align}
\omega=\pm\,c_s\,\sqrt{q_1^2+q_2^2}\,,\quad \omega=0\,,
\end{align}
with the speed of sound $c_s^2=s/(T\partial_T S)$. The relations \eqref{eq:eoms_ideal} are going to be useful later, in section \ref{sec:pert_hydro} when we derive the dissipative part of the stress tensor components of equation \eqref{eq:const_rels}.

\section{The Constitutive Relations}\label{sec:const_relations}

In this section we will extract the next to leading order correction for the VEVs of the hydrodynamic perturbations we started constructing in section \ref{sec:pert}. In order to achieve this, we will use the technique surrounding the asymptotic behaviour of the radial component of the symplectic current in equation \eqref{eq:scurrent_asymptotics}.

To make our strategy more specific, we will consider the symplectic current for the hydrodynamic ansatz \eqref{eq:hydro_pert_exp} with the static solutions summarised by equations \eqref{eq:static_pert_dT}, \eqref{eq:static_pert_dv} and \eqref{eq:static_pert_source}. The next step is to examine the implications of the conservation equation \eqref{eq:div_free} order by order in an $\varepsilon$ expansion.

At leading order in $\varepsilon$, it is only the radial component of the symplectic current that will carry non-trivial information since any derivative along the boundary directions will introduce additional factors of $\varepsilon$. Moreover, at this order, only the term $\delta f^{(1)}$ of our mode will contribute to the symplectic current. Since $\delta f^{(1)}$ is itself a linear combination of the static solutions \eqref{eq:static_pert_dT}, \eqref{eq:static_pert_dv} and \eqref{eq:static_pert_source}, it is obvious that this should be trivially satisfied since the static perturbations are solutions themselves. We list the corresponding conditions in Appendix \ref{app:static_constraints} as these will be useful in subsection \ref{sec:stress_vs_hydro}.

In subsection \ref{sec:stress_vs_horizon} we will use the static perturbations generated by the large diffeomorphisms of section \ref{sec:pert_diff}. This will allow us to relate the VEVs of the stress tensor components, at order $\varepsilon^2$, to horizon data of the perturbation and bulk integrals over the static solutions. These integrals will come from integrating the terms in \eqref{eq:div_free} containing partial derivatives along the boundary directions. 

The final two steps will be made in section \ref{sec:stress_vs_hydro} where we will consider the symplectic current formed by our hydrodynamic perturbation \eqref{eq:hydro_pert_exp} and the static solutions of section \ref{sec:pert_therm}. This will allow us to express the horizon data that will appear in the expressions for the stress tensor components in section \ref{sec:stress_vs_horizon}, in terms of the hydrodynamic variables $\delta T$ and $\delta v^i$. However, there will still be bulk integrals which we will simplify using the constraints of Appendix \ref{app:static_constraints}. This will leave us with simpler bulk integrals which are pure artefacts of the fluid frame. We will remove them by moving our description to the Landau frame where all bulk integrals will eventually drop out.

\subsection{The Stress Tensor in Terms of Horizon Data}\label{sec:stress_vs_horizon}

In this section we will consider the sympectic current $P^\mu_{\delta s_{ab},\delta H}$ that we can form based on our hydrodynamic expansion \eqref{eq:hydro_pert_exp} and the static solutions of equation \eqref{eq:static_pert_source}. Our first step is to consider the expansion of the current in $\varepsilon$ according to,
\begin{align}\label{eq:sympl_curr_eexp}
P^r_{\delta s_{ab},\delta H}&=\delta s_{ab}\,\varepsilon\,P^{r(1)}_{\delta_{s_{ab}},\delta H}+\delta s_{ab}\,\varepsilon^2\,P^{r(2)}_{\delta_{s_{ab}},\delta H}+\cdots\,,\nn
P^c_{\delta s_{ab},\delta H}&=\delta s_{ab}\,\varepsilon\,P^{c(1)}_{\delta_{s_{ab}},\delta H}+\cdots\,.
\end{align}
As we explained in the beginning of the section, the only term of the hydrodynamic expansion \eqref{eq:hydro_pert_exp} that contributes to $P^{r(1)}_{\delta s_{ab},\delta H}$, is the term $\delta f^{(1)}$ which is a linear combination of the static solutions. Therefore, the constraints we list in Appendix \ref{app:static_constraints} will make sure that the leading part,
\begin{align}
\partial_r P^{r(1)}_{\delta_{s_{ab}},\delta H}=0\,,
\end{align}
of the divergence free condition \eqref{eq:div_free} is satisfied.

Moving on to the next order in $\varepsilon$, we see that the terms appearing in \eqref{eq:sympl_curr_eexp} have to satisfy,
\begin{align}
\partial_r P^{r(2)}_{\delta_{s_{ab}},\delta H}-i\omega\,\left(P^{t(1)}_{\delta_{s_{ab}},\delta H}+S^{\prime}\,P^{r(1)}_{\delta_{s_{ab}},\delta H}\right)+iq_i\,P^{i(1)}_{\delta_{s_{ab}},\delta H}=0\,.
\end{align}
After integrating this equation in the bulk from the horizon to the conformal boundary we obtain,
\begin{align}
\delta\langle T^{ab} \rangle_{(2)}&=\left.P^{r(2)}_{\delta_{s_{ab}},\delta H}\right|_{r=0}+i\,\int_0^\infty dr\,\left( \omega\,\left(P^{t(1)}_{\delta_{s_{ab}},\delta H}+S^{\prime}\,P^{r(1)}_{\delta_{s_{ab}},\delta H}\right)-q_i\,P^{i(1)}_{\delta_{s_{ab}},\delta H}\right)\nn
&=\left.P^{r(2)}_{\delta_{s_{ab}},\delta H}\right|_{r=0}+B^{ab}\,,
\end{align}
which also defines the quantity $B^{ab}$. To obtain it we used our observation in equation \eqref{eq:scurrent_asymptotics} regarding the asymptotics of the symplectic current. From the above equation it is clear that the leading correction to the VEV of the stress tensor will come from evaluating the radial component of the symplectic current on the black hole horizon. We see that the bulk integrals that come from integrating the divergence free condition are not zero.

After explicit evaluation we obtain,
\begin{align}\label{eq:stress_tensor_horizon}
\varepsilon\,\delta\langle T^{tt} \rangle_{(2)}&=\frac{i}{4\pi}\left(-s\,q_i\,\left(\delta s^{it}+\delta v^i \right)+\omega\,\left(s\,\delta^{ij}\,\delta s_{ij}+\partial_T \,s\,\delta T\right)-\varepsilon\,i\,8\pi^2\,T\,\delta^{ij}\,\delta g^{(2)(0)}_{ij} \right)+\varepsilon\,B^{tt}\,,\nn
\varepsilon\,\delta\langle T^{ti} \rangle_{(2)}&=-\frac{i}{4\pi}\left( q^i\,\left(s\,\delta^{kl}\,\delta s_{kl}+\partial_T s\,\delta T\right)+\omega\,s\,\delta v^i+\varepsilon\,i\,s\,\delta^{ij} \left(\delta g_{tj}^{(2)(1)}-2\,\delta g_{tj}^{(2)(0)}\,g_{(1)} \right)\right)+\varepsilon\,B^{ti}\,,\nn
\delta\langle T^{ti} \rangle_{(2)}&=-4\pi\,T\,\delta^{ij}\,\delta g_{tj}^{(2)(0)}+B^{it}\,,\nn
\varepsilon\,\delta\langle T^{ij} \rangle_{(2)}&=-\frac{is}{4\pi}\left(2\,\left(q^{\left(i\right.}\delta s^{\left.j\right)t}+q^{\left(i\right.}\delta v^{\left.j\right)} \right)-\delta^{ij}\, \left(q_k\, \delta s^{kt}+q_k\,\delta v^k \right)\right)-\varepsilon\,\frac{s T}{2}\delta^{ij}\,\left(\delta g_{rr}^{(2)(0)}+\delta g_{tt}^{(2)(0)} \right)\nn
&\qquad-\frac{i\,\omega}{4\pi}\left(\delta^{ij}\,\left(2s\,\delta^{kl}\,\delta s_{kl}+\partial_T s\,\delta T-s\,\delta s_{tt} \right)-2s\,\delta s^{(ij)}\right)+\varepsilon\,B^{ij}\,.
\end{align}
It is interesting to note that we can obtain two inequivalent expressions for the $ti$ components of the stress tensor, depending on the static solution we use to read it off. However, the two static solutions are related by a perturbative boost which is also a solution to the equations of motion.

As we can see more explicitly, equations \eqref{eq:stress_tensor_horizon} express the stress tensor components in terms of horizon data plus a bulk integral. However, we have not yet achieved to express it solely in terms of the local temperature $\delta T$ and fluid velocity $\delta v^i$ fluctuations. Instead, we have the appearance of the constants of integration $\delta g_{ij}^{(2)(0)}$,  $\delta g_{ti}^{(2)(0)}$, $\delta g_{rr}^{(2)(0)}$ and $\delta g_{tt}^{(2)(0)}$ as well as the expansion coefficient $\delta g_{ti}^{(2)(1)}$. This is the first aim of the next subsection where we will manage to eliminate them from the constitutive relations. As we will see, the constants of integration $\delta g_{ti}^{(2)(0)}$ and the combination $\delta g_{rr}^{(2)(0)}+\delta g_{tt}^{(2)(0)}$ can be simply removed by redefining $\delta v^i$ and $\delta T$. Another way to see this is to observe that adding the static solutions to the correction $\delta f^{(2)}$ would still give a solution at that order in $\varepsilon$. Those constants of integration can therefore be thought of as the corrections of the local temperature and fluid velocity at order $\varepsilon^2$.

\subsection{The Stess Tensor in Terms of Hydrodynamic Variables}\label{sec:stress_vs_hydro}
In this section we will finalise the derivation of our hydrodynamics and we will show that it indeed agrees with the general intuition of relativistic hydrodynamics. However, during the process we will derive expressions for the shear and bulk viscosities $\eta$ and $\zeta$ in terms of the black hole horizon data of equation \eqref{eq:nh_bcs}.

Our first step is to eliminate the constants of integration from the expression \eqref{eq:stress_tensor_horizon} by expressing them in terms of derivatives our hydrodynamic variables $\delta T$ and $\delta v^i$. In order to achieve this, we consider the symplectic current \eqref{eq:scurrent_def} when the role of the static perturbation is played by one of the solutions obtained from themodynamic variations. These static perturbations were discussed in section \ref{sec:pert_hydro} and we will call $P^\mu_{\delta T,\delta H}$ and $P^\mu_{\delta v^i,\delta H}$ the symplectic current corresponding temperature and fluid velocity variations correspondingly.

Similarly to the analysis of section \ref{sec:stress_vs_horizon}, after expanding the symplectic currents in $\varepsilon$, the divergence free condition \eqref{eq:div_free} yields the non-trivial equations,
\begin{align}
\partial_r P^{r(2)}_{\delta_T,\delta H}-i\omega\,\left(P^{t(1)}_{\delta_T,\delta H}+S^{\prime}\,P^{r(1)}_{\delta_T,\delta H}\right)+iq_i\,P^{i(1)}_{\delta_T,\delta H}=&0\,,\nn
\partial_r P^{r(2)}_{\delta_{v^j},\delta H}-i\omega\,\left(P^{t(1)}_{\delta_{v^j},\delta H}+S^{\prime}\,P^{r(1)}_{\delta_{v^j},\delta H}\right)+iq_i\,P^{i(1)}_{\delta_{v^j},\delta H}=&0\,.
\end{align}
After integrating from the horizon up to the conformal boundary we obtain the relation,
\begin{align}
\left.P^{r(2)}_{\delta_T,\delta H}\right|_{r=0}+B_T=&0\,,\nn
\left.P^{r(2)}_{\delta_{v^j},\delta H}\right|_{r=0}+B_j=&0\,.
\end{align}
To obtain the above equations we used that both our static solutions have zero sources at infinity. This implies that, according to equation \eqref{eq:scurrent_asymptotics}, the radial component of the symplectic current must asymptote to zero at infinity. For clarity we write the defining equations,
\begin{align}
B_T&=i\,\int_0^\infty dr\,\left( \omega\,\left(P^{t(1)}_{\delta_{T},\delta H}+S^{\prime}\,P^{r(1)}_{\delta_{T},\delta H}\right)-q_i\,P^{i(1)}_{\delta_{T},\delta H}\right)\,,\nn
B_j&=i\,\int_0^\infty dr\,\left( \omega\,\left(P^{t(1)}_{\delta_{v^j},\delta H}+S^{\prime}\,P^{r(1)}_{\delta_{v^j},\delta H}\right)-q_i\,P^{i(1)}_{\delta_{v^j},\delta H}\right)\,,
\end{align}
with explicit expressions in terms of bulk fields appearing in Appendix \ref{app:bulk_integrals}.

Evaluating the radial components on the horizon by using the boundary conditions \eqref{eq:nh_bcs} and \eqref{eq:gen_exp}, we obtain the relations,
\begin{align}\label{eq:dTdV_relations}
-\frac{(\partial_T s)^2}{s}\omega\,\delta T+\varepsilon\,8i\pi^2\, \delta^{ij}\delta g_{ij}^{(2)(0)}-\omega\,\partial_T s\,\left(\,\delta^{ij}\,\delta s_{ij}-\delta s_{tt}\right)\qquad\qquad&\nn
\quad +\varepsilon\,2i\,\pi\,T\partial_T s\left( \delta g_{rr}^{(2)(0)}+\delta g_{tt}^{(2)(0)}\right)+s\,\omega\,\Phi_{IJ}^{(0)}\,\partial_T\varphi^{I(0)}\,\partial_T\varphi^{J(0)}\,\delta T&=-\varepsilon\,4\pi i B_T\,,\nn
i\,q_j\left(s\,\delta^{kl}\delta s_{kl} \partial_T+ s\,\delta T\right)-\varepsilon\,2\,\left( 8\pi^2\,T-s\,g_{(1)}\right)\delta g_{tj}^{(2)(0)}+s\left( i\omega\,\delta v_j-\delta g_{tj}^{(2)(1)}\right)&=\varepsilon\,4\pi B_j\,.
\end{align}
The second equation together with the relation,
\begin{align}
B^{ti}-B^{it}=-\delta^{ij}B_j\,,
\end{align}
between the bulk integrals, simply implies the equivalence between the two expressions for the $ti$ component of the stress tensor in equation \eqref{eq:stress_tensor_horizon}. The first equation in \eqref{eq:dTdV_relations} can be used to eliminate the constants of integration $\delta g_{ij}^{(2)(0)}$ from the $tt$ component of equation \eqref{eq:stress_tensor_horizon} to yield,
\begin{align}
\varepsilon\,\delta\langle T^{tt} \rangle_{(2)}=&-\varepsilon\,\frac{1}{2}\partial_T s\,T^2\,\left(\delta g_{rr}^{(2)(0)}+\delta g_{tt}^{(2)(0)} \right)+i\omega\,\frac{1}{4\pi}T\,\partial_T s\,\delta s_{tt}\nn
&\,+i\omega\,\frac{s}{4\pi}\left(\frac{\partial_T s}{s}-\frac{(\partial_T s)^2}{s^2}T+T\,\Phi_{IJ}^{(0)}\,\partial_T\varphi^{I(0)}\,\partial_T\varphi^{J(0)}\right)\,\delta T\nn
&\,+i\omega\,\frac{s}{4\pi}\left(1-T\,\frac{\partial_T s}{s} \right)\,\delta^{ij}\delta s_{ij}-\frac{is}{4\pi}q_i\left(\delta s^{it}+\delta v^{i} \right)\nn
&\,-\varepsilon\,T\,B_T+\varepsilon\,B^{tt}\,.
\end{align}

Using the ideal order fluid equations \eqref{eq:eoms_ideal}, we can show the important identity
\begin{align}\label{eq:bulk_int_identities}
B^{ij}=\delta^{ij}\frac{s}{T\,\partial_T S}\left(-T\,B_T +B^{tt} \right)\,,
\end{align}
for the bulk integrals. The above relation shows that after performing the change of fluid frame,
\begin{align}
\delta T\to \delta T-\frac{\varepsilon}{T\,\partial_T s}\,B^{tt}\,,\qquad \delta v^i\to \delta v^i-\frac{\varepsilon}{\epsilon+p}\,B^{ti}\,,
\end{align}
we can eliminate the bulk integrals from our constitutive relations. At this point, this might seem like a miracle. However, by observing the form of the bulk integrals that we list in Appendix \ref{app:bulk_integrals}, the integrands are all proportional to derivatives of the function $S$ which was chosen arbitrarily in equation \eqref{eq:fourier_modes} when we wrote the form of our Fourier modes. All that we can fix about this function is its leading behaviour close to the black hole horizon at $r=0$ and that it has to vanish sufficiently fast at the conformal infinity. After fully fixing the fluid frame, the transport coefficients in our constitutive relations cannot depend on coordinate system dependent quantities in the bulk and the single remaining non-trivial combination of the bulk integrals has to vanish.

Taking a step further, we perform the change of hydrodynamic variables,
\begin{align}
\delta T\to \delta T-\frac{\varepsilon}{T\,\partial_T s}\,\delta\langle T^{tt} \rangle_{(2)}\,,\qquad \delta v^i\to \delta v^i-\frac{\varepsilon}{\epsilon+p}\,\delta\langle T^{ti} \rangle_{(2)}\,,
\end{align}
along with the fluid equations \eqref{eq:eoms_ideal}, to obtain the constitutive relations in the Landau frame,
\begin{align}\label{eq:landau_frame}
\varepsilon^2\,\delta\langle T^{ij} \rangle_{(2)}=&\eta\,\delta^{ik}\delta^{jl}\left( \delta_{kl}\,\delta^{rs}\left(\partial_t\delta s_{rs}-\partial_r\delta s_{st}\right) -2\,\left(\partial_t\delta s_{\left(kl\right)} -\partial_{\left(k\right.}\delta s_{\left.l\right)t}\right)\right)\nn
&\quad +\eta\,\delta^{ik}\delta^{jl}\left( \delta_{kl}\,\delta^{rs}\partial_r\delta v_s -2\,\partial_{\left(k\right.}\delta v_{\left.l\right)} \right)+\zeta\,\delta^{ij}\,\delta^{kl}\,\left( \partial_k\delta s_{lt}-\partial_t\delta s_{kl}-\partial_k\delta v_l\right)\,,\nn
\delta\langle T^{ti} \rangle_{(2)}=&0\,,\quad \delta\langle T^{tt} \rangle_{(2)}=0\,.
\end{align}
In the above we have defined the transport coefficients,
\begin{align}\label{eq:eta_zeta}
\eta=\frac{s}{4\pi}\,,\quad \zeta=\frac{s}{4\pi}\left(\frac{s}{\partial_T s} \right)^2\,\Phi_{IJ}^{(0)}\,\partial_T\varphi^{I(0)}\,\partial_T\varphi^{J(0)}\,,
\end{align}
with $\eta$ and $\zeta$ being the shear and the bulk viscosities respectively.

We would now like to show that the identification of equation \eqref{eq:eta_zeta} is indeed correct. For this reason, we remind the reader that the leading dissipative part of the stress tensor in the Landau frame reads,
\begin{align}
\tau^{\mu\nu}&=-\eta\,\Delta^{\mu\alpha}\Delta^{\nu\beta}\left(2\,\nabla_{\left(\alpha\right.}u_{\left.\beta\right)}-g_{\alpha\beta}\,\nabla_\lambda u^\lambda \right)-\zeta\,\Delta^{\mu\nu}\,\nabla_\lambda u^\lambda\,,\nn
\Delta^{\mu\nu}&=g^{\mu\nu}+u^\mu\,u^\nu\,.
\end{align}
One can easily show that this agrees with the expressions of \eqref{eq:landau_frame} given that,
\begin{align}
ds^2&=g_{\mu\nu}\,dx^\mu\,dx^\nu=\left(\eta_{\mu\nu}+2\,\delta s_{\left(\mu\nu\right)}\right)\,dx^\mu\,dx^\nu\nn
&=\eta_{\mu\nu}\,dx^\mu\,dx^\nu+\delta s_{it}\,dt\,dx^i+2\,\delta s_{\left(ij\right)}\,dx^i\,dx^j\,,\nn
u&=\left(1+\delta s^{tt}\right)\,\partial_t+\left(\delta v^i+\delta s^{it}\right)\,\partial_i\,,
\end{align}
along with the matching condition \eqref{eq:eta_zeta}.

\section{Numerical Checks}\label{sec:numerics}
In this section we perform numerical checks on the bulk viscosity result of the previous section. In order to do this, we focus on the model \eqref{eq:bulk_action} with $I,J=1,2$, where 

\begin{align}\label{model}
&V(\varphi^1, \varphi^2)=-6 -(\varphi^1)^2+\lambda_1(\varphi^1)^4-(\varphi^2)^2+\lambda_3(\varphi^2)^4-\lambda_2(\varphi^1)^2(\varphi^2)^2\,,\nonumber\\
&\Phi_{IJ}(\varphi^1, \varphi^2)=\frac{\delta_{IJ}}{2}(\cosh(\varphi^1)+\cosh(\varphi^2))\,.
\end{align}
Note that the scalars correspond to operators in the dual field theory with scaling dimensions $\Delta_1=\Delta_2=2$. In what follows we will focus our attention on the case $\lambda_1=\lambda_3=1\,, \lambda_2=0\,$.

Given the ansatz \eqref{eq:background} and the UV and IR boundary expansions, \eqref{eq:uv_bcs} and \eqref{eq:nh_bcs} respectively, we proceed to solve the boundary condition problem numerically for boundary deformations for the scalars given by $\varphi^1_{(s)}=1, \varphi^2_{(s)}=1/2$. This configuration corresponds to our background solution, around which we will consider perturbations in order to compute the bulk viscosity. 

\subsection{Quasinormal modes}
We now move on to compute quasinormal modes for the backgrounds constructed above. We consider perturbations of the form
\begin{align}
\label{eq:deltag}
&\delta ds^2= - U \delta h_{tt} dt^2+2U \delta h_{t\,x_i}dt dx_i+e^{2 g} \left( h_{11} dx_1^2+h_{22} dx_2^2+2 h_{12}dx_1 dx_2\right)\,,
\end{align}
together with $\delta \varphi^1, \delta \varphi^2$,
where the variations are taken to have the form
\begin{equation}
f(t,r,x_1)=e^{-i \omega v(t,r)+i q x_1}f(r)\,,
\end{equation}
with $v$ the Eddington-Finkelstein coordinate defined as
\begin{align}
v(t,r,x_1)=t+\int_{\infty}^{r}\frac{dy}{U(y)}\,.
\end{align}
Note that our choice for the momentum $q$ to point in the direction $x_1$ is without loss of generality, because the background is isotropic. Plugging this ansatz in the equations of motion, we obtain 4 first order ODEs and 4 second order giving rise to 12 integration constants.

We now outline the boundary conditions for these fields. In the IR, we impose in-falling boundary conditions at the horizon
\begin{alignat}{2}
&\delta h_{tt}=c_1 \,r+\dots\,,\qquad&&\nonumber\\
&\delta h_{t \,x_1}=c_2 +\dots\,,\qquad&&\delta h_{t \,x_2}=c_3 +\dots\,,\nonumber\\
&\delta h_{x_1x_1}=c_4+\dots\,,\qquad&&\delta h_{x_2 x_2}=-c_4+\dots\,,\nonumber\\
&\delta h_{x_1 \,x_2}=c_5 +\dots\,,\qquad&&\delta\varphi^1=c_6+\dots\,,\nonumber\\
&\delta\varphi^2=c_{7}+\dots\,,&&
\end{alignat}
where the constants $c_1,c_2,c_3$ are not free but are fixed in terms of the others. Thus, for fixed value of $q$, the expansion is fixed in terms of 5 constants, $\omega,c_4,c_5,c_6,c_7$. 

On the other side of the geometry, in UV, one can write down the following expansion 
\begin{alignat}{2}
&\delta h_{tt}=\delta h_{tt}^{(s)}+\dots\,,&&\nonumber\\
&\delta h_{t x_1}=\delta h_{t\,x_1}^{(s)}+\dots\,,\qquad&&\delta h_{t x_2}=\delta h_{t\,x_2}^{(s)}+\dots,,\nonumber\\
&\delta h_{x_1x_1}=\delta h_{x_1\,x_1}^{(s)}+\dots\,,\qquad&&\delta h_{x_2 x_2}=\delta h_{x_2\,x_2}^{(s)}+\dots+\frac{\delta h_{x_2\,x_2}^{(v)}}{(r+R)^3}+\dots\,,\nonumber\\
&\delta h_{x_1\, x_2}=\delta h_{x_1\,x_2}^{(s)}+\dots+\frac{\delta h_{x_1\,x2}^{(v)}}{(r+R)^3}+\dots\,, 
&&\delta\varphi^1=\frac{\delta\varphi^1_{(s)}}{(r+R)}+\frac{\delta\varphi^1_{(v)}}{(r+R)^2}+\dots\,,\nonumber\\
&\delta\varphi^2=\frac{\delta\varphi^2_{(s)}}{(r+R)}+\frac{\delta\varphi^2_{(v)}}{(r+R)^2}+\dots\,&&
\label{eq:pertUV}
\end{alignat}
For the computation of quasinormal modes, we need to ensure that we remove all the sources from the UV expansion up to a combination of coordinate reparametrisations 
\begin{align}\label{eq:coord_transf}
[\delta g_{\mu\nu}+\mathcal{L}_{\tilde \zeta} g_{\mu\nu}]\to 0\,,
\end{align} 
where the transformations are of the form
\begin{align}\label{eq:coord_transf}
x^\mu\to x^\mu+\tilde\zeta^\mu\,\quad\tilde \zeta&=e^{-i\omega t+i q x_1}\,\tilde\zeta^\mu \,\partial_\mu\,,
\end{align}
for $\zeta$ constant. This requirement boils down to the sources appearing in \eqref{eq:pertUV} taking the form
 \begin{align}
 \delta h_{t t}^{(s)}&=2 i \omega\, \tilde\zeta_1-2\tilde \zeta_2 \,,\notag\\
 \delta h_{t x_{1}}^{(s)}&=iq\,\tilde \zeta_1+i \omega \,\tilde\zeta_3\,,\notag\\
  \delta h_{t x_{2}}^{(s)}&=i \omega \,\tilde\zeta_4\,,\notag\\
 \delta h_{x_1\, x_{1}}^{(s)}&=-2 \tilde\zeta_2-2 i q \,\tilde\zeta_3\,,\notag\\
 \delta h_{x_2\, x_{2}}^{(s)}&=-2 \tilde\zeta_2\,,\notag\\
  \delta h_{x_1\, x_{2}}^{(s)}&=-i q\tilde \zeta_4\,,\notag\\
 \delta\varphi^1_{(s)}&=0\,,\notag\\ 
  \delta\varphi^2_{(s)}&=0\,.
   \end{align}
 We now see that the UV expansion is fixed in terms of 8 constants: $\tilde\zeta_1,\tilde\zeta_2,\tilde\zeta_3, \tilde\zeta_4$ and $\delta h_{x_2\,x_2}^{(v)},\delta h_{x_1\,x_2}^{(v)}, \delta\varphi^1_{(v)},\delta\varphi^2_{(v)}$. Overall, for fixed $q$, we have 13 undetermined constants, of which one can be set to unity because of the linearity of the equations. This matches precisely the 12 integration constants of the problem and thus we expect our solutions to be labelled by $q$.

We proceed to solve numerically this system of equations subject to the above boundary conditions using a double-sided shooting method. We expect to find a pair of sound modes with dispersion relation
\begin{equation}
\omega=\pm q \, c_s -i\frac{\Gamma}{2}\, q^2\,,
\end{equation}
where the attenuation $\Gamma$ is given in terms of the bulk viscosity $\zeta$ through
\begin{equation}
\label{eq:attenuation}
\Gamma=\frac{1}{s \,T} \left( \frac{s}{4\pi} +\zeta\right) 
\end{equation}
and $c_s^2=s/(T\partial_T S)$ is simply the speed of sound.

\begin{figure}[h!]
\centering
\includegraphics[width=0.45\linewidth]{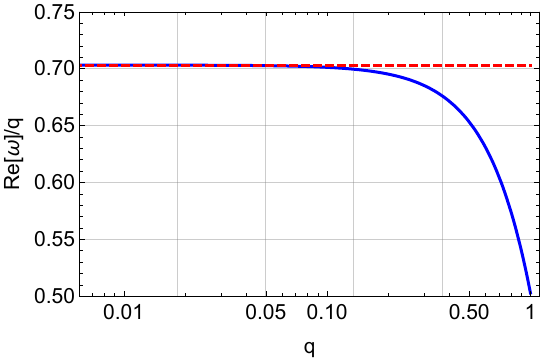}\quad \includegraphics[width=0.45\linewidth]{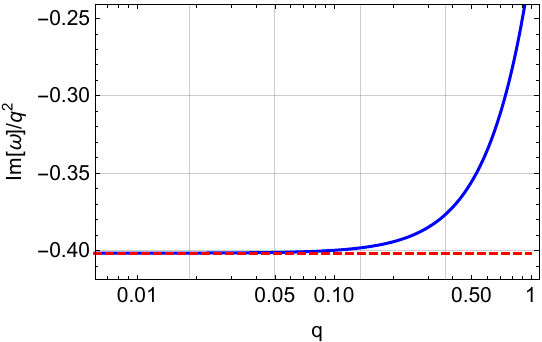}
\caption{The solid lines correspond to $Re[\omega]/q$ and $Im[\omega]/q^2$ as functions of $q$ in the hydrodynamic limit. The dashed lines correspond to $c_s$ and $\Gamma/2$, calculated using the analytic expressions \eqref{eq:attenuation} and \eqref{eq:eta_zeta}. Here $\lambda_1=\lambda_3=1,\lambda_2=0, \varphi^1_{(s)}=1, \varphi^2_{(s)}=1/2, T=1/10$.}
\label{fig:qnms}
\end{figure}

In figure \ref{fig:qnms}, we plot the real and imaginary part of the dispersion relation obtained numerically divided by the appropriate powers of the momentum so that they asymptote to a constant in the hydrodynamic limit. The dashed lines correspond to $c_s$ and $\Gamma/2$, where the latter was calculated using the analytic expression of the bulk viscosity \eqref{eq:eta_zeta}. We see good quantitative agreement.

\section{Summary and Discussion}\label{sec:discussion}

In this paper we presented a general technique to derive the effective hydrodynamic description of long wavelength perturbations of relativistic holographic fluids. The most important ingredient in our construction was the existence of a generalisation of Liouville's theorem for classical gravitational and gauge theories. This is essentially captured by the fact that the Crnkovic-Witten symplectic current is divergence free when evaluated on-shell.

It is worth comparing and contrasting our work with the standard fluid-gravity correspondence. As one would expect, the common ground of both approaches is that the gravitational and gauge constraints in the bulk serve as the conservation laws of the effective theory. The input of the gravitational side is the radial equations which fix the constitutive relations for the stress tensor and the electric currents in terms of the hydrodynamic variables. Our work has made significant progress by showing that explicit solutions for the gravitational equations are not required for this task. We achieved this by reformulating the problem in a way that allowed us to derive the constitutive relations based on general properties of the background black holes. However, we have achieved this only at a linearised level in the hydrodynamic fluctuations. An important question would be to better understand how the non-linearities of hydrodynamics could be dealt with in this framework.

An interesting direction is to apply our techniques in different scenarios as this will provide the extraction of new transport coefficients. A natural example would be superfluids at finite chemical potential which have more transport coefficients to be fixed \cite{Herzog:2011ec,Bhattacharya:2011eea}. In our previous work \cite{Donos:2021pkk}, we only fixed the transport coefficients of the current sector of holographic superfluids at zero chemical potential. However, we expect our technique to be able to fix all transport coefficients of holographic superfluids even at finite chemical potential. This would necessarily include the stress tensor since it couples to the electric current at finite density.

Another important application for our techniques is related to critical phenomena and phase transitions. From the formula \eqref{eq:eta_zeta} for the bulk viscosity, it is easy to conclude that this blows up like $(T_c-T)^{-1}$ close to the critical temperature $T_c$. Earlier numerical studies have revealed similar behaviour of the bulk viscosity near the critical point \cite{Buchel:2009mf}. This is very similar to the situation that was noticed in \cite{Donos:2021pkk} in the context of superfluid phase transitions at zero chemical potential or in the probe limit at finite chemical potential \cite{Herzog:2011ec}. This kind of infinity is related to the fact that we have not included the relevant amplitude mode which becomes gapless at the critical point.  It would therefore be interesting to extend our work on the amplitude mode in order to include it in the effective description of large but finite wavelength hydrodynamic modes.

\section*{Acknowledgements}
We would like to thank V. Ziogas for discussions and collaboration in related topics. AD is supported by STFC grant ST/T000708/1. C.P. acknowledges support from a Royal Society - Science Foundation Ireland University Research
Fellowship via grant URF/R1/211027.

\appendix

\section{The first order action}\label{app:fo_action}
In this Appendix we will examine the sum of the bulk action \eqref{eq:bulk_action} and the Gibbons-Hawking term in the boundary action \eqref{eq:bdy_action}. In particular, we will examine the result of the integration by parts in the bulk leading to an action which only contains first order partial derivatives of the metric components. This will prove the claim that the first order form \eqref{eq:first_order_action} for the total action is equivalent to the sum of \eqref{eq:bulk_action} and \eqref{eq:bdy_action}, provided the asymptotic boundary conditions of equation \eqref{eq:uv_bcs}.
\begin{align}
S_{EH}&=\int_M d^dx\,\sqrt{-g}\,R\nn
&=\int_M d^dx\, \partial_{\kappa}\left(\sqrt{-g}g^{\mu\nu}\Gamma^{\kappa}_{\mu\nu}-\sqrt{-g}g^{\mu\kappa}\Gamma^{\nu}_{\nu\mu}\right)+\int_M d^dx\,\mathcal{L}(g_{\mu\nu},\partial_k g_{\mu\nu})\nn
&=\int_{\partial M}d^{d-1}x\, \sqrt{-h}\,n_r\,\left((g^{\mu\nu}\Gamma^{r}_{\mu\nu}-g^{\mu r}\Gamma^{\nu}_{\nu\mu}\right)+\int_M d^dx\,\mathcal{L}(g_{\mu\nu},\partial_k g_{\mu\nu})\,,
\end{align}
where $\partial M$ is the (conformal) boundary hypersurface defined by a constant value of the radial coordinate $r$ and $n=N\, dr$ the unit norm normal one-form. The induced metric on the hypersurface $\partial M$ is $h_{\mu\nu}=g_{\mu\nu}-n_\mu n_\nu$ and we can decompose our bulk metric according to,
\begin{align}
ds^2=N^2\,dr^2+\gamma_{ac}\,\left(dx^a+N^a\,dr \right)\,\left(dx^c+N^c\,dr \right)
\end{align}

The Gibbons-Hawking term is designed to cancel out the normal derivatives of the variations of the metric coming from the bulk action,
\begin{align}
S_{GH}=2\int_{r\to\infty}d^{d-1}x\, \sqrt{-h}\,K\,,
\end{align}
where $K_{\mu\nu}=h_{\mu}{}^\lambda\,\nabla_\lambda n_{\nu}$ is the extrinsic curature tensor and $K=g^{\mu\nu}\,K_{\mu\nu}$ is its trace. We can now write the Gibbons-Hawking term as,
\begin{align}
S_{GH}=\int_{\partial M}d^{d-1}x\, \sqrt{-h}\,\left(-g^{\mu\nu}\Gamma^{\rho}_{\mu\nu}n_\rho+\Gamma^\mu_{\mu \sigma}g^{\sigma \rho}n_\rho+g^{\mu \rho}\partial_\mu n_\rho+\partial_\mu n^\mu\right)\,.
\end{align}
Adding up the bulk action with the boundary term we obtain,
\begin{align}\label{eq:first_order_taction}
S_{EH}+S_{GH}&=\int_M d^dx\,\mathcal{L}(g_{\mu\nu},\partial_k g_{\mu\nu})+\int_{\partial M}d^{d-1}x \sqrt{-h}\,\left(g^{\mu \rho}\,\partial_\mu n_\rho+\partial_\mu n^\mu\right)\nn
&=\int_M d^dx\,\mathcal{L}(g_{\mu\nu},\partial_k g_{\mu\nu})-\int_{\partial M}d^{d-1}x \sqrt{-h}\,N^{-1}\,\partial_a \left(N^a \right)\,,
\end{align}
where we use that $N^\mu\partial_\mu=\partial_r-N\,n^\mu\,\partial_\mu$. As we can see, even though the total action is covariant, the two individual terms are not. The cause of this lies in the fact that we performed an integration by parts in the second derivative parts of the Ricci tensor and we showed that the resulting bulk term is covariant up to surface terms which are meant to be cancelled by the final surface term.

Strictly speaking, the surface term of the action \eqref{eq:first_order_taction} is meant to be kept at all times in order to guarantee diffeomorphism invariance of the theory. However, it will not contribute to the boundary stress tensor when varying with respect to the boundary metric $\gamma_{ac}$ provided that the shift vector $N^a$ decays fast enough and this can always be guaranteed by an appropriate choice of the bulk coordinate system.

\section{Bulk integrals}\label{app:bulk_integrals}
In this appendix we list the bulk integrals that we obtained after integrating the divergence free condition \eqref{eq:div_free} over the bulk.
\begin{align}
\varepsilon\,B^{tt}&=\int_0^\infty dr \Big( \left(i q_k \left(\delta s^{kt}+\delta v^k\right)-i\omega\,\delta_{kl}\, \delta s^{kl} \right)\left(e^{2g}\,U S^\prime\right)^\prime- 2 i\omega \,\delta T \left(e^{2g} U S^\prime \partial_{T}g\right)^\prime \Big)\,,\nn\nn
\varepsilon\,B^{ij}&=\delta^{ij}\,\int_0^\infty dr \Big( \left(-i q_k \left(\delta s^{kt}+\delta v^k \right)+i\omega\, \delta s^{tt}\right)\left(e^{2g}\,U S^\prime\right)^\prime\nn
&\quad +i\omega\,\delta T\left(-\, \partial_{T}U\, \frac{e^{2g}}{U}\left(U S^\prime\right)^\prime+2 e^{2g}U S^\prime\partial_T g^\prime+ U e^{2g}\, S^\prime\,\partial_T \phi^{I} \,\Phi_{IJ}\, \phi^{J \prime}\right)\Big)\,,\nn\nn
\varepsilon\,B^{it}&=i q^i\,\int_0^\infty dr\,\Big(  \left(-\delta_{kl} \,\delta s^{kl}+\delta s^{tt}-2 \delta T\, \partial_T g\right)\left(e^{2g}U S^\prime \right)^\prime\nn
&\qquad-\frac{e^{2g}}{U}\delta T \left(\partial_T U\left(S^\prime U\right)^\prime-U^2 S^\prime \partial_T \phi^{I}\, \Phi_{IJ}\, \phi^{J \prime}\right)\Big)\,,\nn\nn
\varepsilon\,B_T&= \int_0^\infty dr \left(   \left(-i q_k (\delta s^{kt}+\delta v^k)+i\omega\,\delta_{kl}\, \delta s^{kl}\right) \left(\frac{e^{2g}}{U} \partial_T U \left(U S^\prime\right)^\prime-\partial_T\phi^{I}\, \Phi_{IJ}\, \phi^{J\prime}\,e^{2g}\,U\, S^\prime\right)\right.\nn
&\left.+2 \left(-i q_k \left(\delta s^{kt}+\delta v^k\right)+i\omega\, \delta s^{tt}\right)\,\partial_T g \left( e^{2g} U S^\prime \right)^\prime +2 i \omega \,\left(-\delta_{kl}\, \delta s^{kl}+\delta s^{tt}\right) e^{2g} U \partial_T g^\prime S^\prime \right)
\end{align}
Moreover, we have the two identities,
\begin{align}
B^{ti}=0\,,\qquad B_i=B^{it}\,,
\end{align}
which can be easily obtained by direct evaluation of the symplectic current. Based on the above expressions, one can show the identities \eqref{eq:bulk_int_identities} by using the ideal fluid equations \eqref{eq:eoms_ideal} and performing integrations by parts in order to get rid of the second derivatives of the function $S$.

\section{Constraints for static perturbations}\label{app:static_constraints}

Here we will discuss the constraints which are implied by the divergence free condition \eqref{eq:div_free} of the symplectic current when considered for pairs of the static perturbative solutions that we discussed in sections \ref{sec:pert_therm} and \ref{sec:pert_diff}. For this, we need to consider all different pais of solutions to form the symplectic currents $P^\mu_{\delta_{s_{\mu\nu}},\delta_{s_{\rho\sigma}}}$, $P^\mu_{\delta_{s_{\mu\nu}},\delta_{v_i}}$, $P^\mu_{\delta_{s_{\mu\nu}},\delta_T}$ and $P^\mu_{\delta_T,\delta_{v_i}}$. In this paper, these constraints have not been used in our derivations but we list them here for completeness.

The solutions that enter all these symplectic currents are independent of the boundary directions and therefore only the radial components of the current will carry non-trivial information. For these, we can therefore write,
\begin{align}
P^r_{\delta_1,\delta_2}(r)=P^r_{\delta_1,\delta_2}(r=0),
\end{align}
yielding the constraints,
\begin{align}\label{eq:constraints}
e^{2g}\,(U^\prime-2\,Ug^\prime)&=T s\,,\nn
-e^{2g}\,(-2\,U^\prime \,\partial_T g +2\,g^\prime\,\partial_T U +4\, U\, \partial_T g^\prime+4\, U\, g^\prime\partial_T g +U\, \partial_T\phi^I\,\Phi_{IJ}\,\phi^{J \prime})&= T\,\partial_T s\,, \nn
e^{2g}\,(\partial_T U^\prime +2\,U\,\partial_Tg^\prime+U\,\partial_T\phi^I\,\Phi_{IJ}\,\phi^{J \prime})&=s\,.
\end{align}
In particular, the first constraint can be found using any of the $P^\mu_{\delta_{s_{tt}},\delta_{s_{ii}}}$,  $P^\mu_{\delta_{s_{ti}},\delta_{v_i}}$, $P^\mu_{\delta_{s_{it}},\delta_{v_i}}$ or $P^\mu_{\delta_{s_{it}},\delta_{s_{ti}}}$. The second using $P^\mu_{\delta_{s_{tt}},\delta_T}$ and the third using $P^\mu_{\delta_T, \delta_{s_{ii}}}$.

\newpage

\bibliographystyle{utphys}
\bibliography{refs}{}

\providecommand{\href}[2]{#2}\begingroup\raggedright\begin{thebibliography}{10}

\bibitem{Hartnoll:2016apf}
S.~A. Hartnoll, A.~Lucas, and S.~Sachdev, ``{Holographic quantum matter},''
\href{http://arxiv.org/abs/1612.07324}{{\ttfamily arXiv:1612.07324 [hep-th]}}.

\bibitem{Kovtun:2012rj}
P.~Kovtun, ``{Lectures on hydrodynamic fluctuations in relativistic
  theories},'' \href{http://dx.doi.org/10.1088/1751-8113/45/47/473001}{{\em J.
  Phys.} {\bfseries A45} (2012) 473001},
\href{http://arxiv.org/abs/1205.5040}{{\ttfamily arXiv:1205.5040 [hep-th]}}.

\bibitem{Aharony:1999ti}
O.~Aharony, S.~S. Gubser, J.~M. Maldacena, H.~Ooguri, and Y.~Oz, ``{Large N
  field theories, string theory and gravity},''
  \href{http://dx.doi.org/10.1016/S0370-1573(99)00083-6}{{\em Phys. Rept.}
  {\bfseries 323} (2000) 183--386},
\href{http://arxiv.org/abs/hep-th/9905111}{{\ttfamily arXiv:hep-th/9905111
  [hep-th]}}.

\bibitem{Witten:1998qj}
E.~Witten, ``{Anti-de Sitter space and holography},''
  \href{http://dx.doi.org/10.4310/ATMP.1998.v2.n2.a2}{{\em Adv. Theor. Math.
  Phys.} {\bfseries 2} (1998) 253--291},
\href{http://arxiv.org/abs/hep-th/9802150}{{\ttfamily arXiv:hep-th/9802150
  [hep-th]}}.

\bibitem{Hartnoll:2008vx}
S.~A. Hartnoll, C.~P. Herzog, and G.~T. Horowitz, ``{Building a Holographic
  Superconductor},''
  \href{http://dx.doi.org/10.1103/PhysRevLett.101.031601}{{\em Phys. Rev.
  Lett.} {\bfseries 101} (2008) 031601},
\href{http://arxiv.org/abs/0803.3295}{{\ttfamily arXiv:0803.3295 [hep-th]}}.

\bibitem{Gubser:2008px}
S.~S. Gubser, ``{Breaking an Abelian gauge symmetry near a black hole
  horizon},'' \href{http://dx.doi.org/10.1103/PhysRevD.78.065034}{{\em Phys.
  Rev.} {\bfseries D78} (2008) 065034},
\href{http://arxiv.org/abs/0801.2977}{{\ttfamily arXiv:0801.2977 [hep-th]}}.

\bibitem{Gubser:2008zu}
S.~S. Gubser, ``{Colorful horizons with charge in anti-de Sitter space},''
  \href{http://dx.doi.org/10.1103/PhysRevLett.101.191601}{{\em Phys.Rev.Lett.}
  {\bfseries 101} (2008) 191601},
\href{http://arxiv.org/abs/0803.3483}{{\ttfamily arXiv:0803.3483 [hep-th]}}.

\bibitem{Nakamura:2009tf}
S.~Nakamura, H.~Ooguri, and C.-S. Park, ``{Gravity Dual of Spatially Modulated
  Phase},'' \href{http://dx.doi.org/10.1103/PhysRevD.81.044018}{{\em Phys.
  Rev.} {\bfseries D81} (2010) 044018},
\href{http://arxiv.org/abs/0911.0679}{{\ttfamily arXiv:0911.0679 [hep-th]}}.

\bibitem{Donos:2011bh}
A.~Donos and J.~P. Gauntlett, ``{Holographic striped phases},''
  \href{http://dx.doi.org/10.1007/JHEP08(2011)140}{{\em JHEP} {\bfseries 08}
  (2011) 140},
\href{http://arxiv.org/abs/1106.2004}{{\ttfamily arXiv:1106.2004 [hep-th]}}.

\bibitem{Horowitz:2012ky}
G.~T. Horowitz, J.~E. Santos, and D.~Tong, ``{Optical Conductivity with
  Holographic Lattices},''
  \href{http://dx.doi.org/10.1007/JHEP07(2012)168}{{\em JHEP} {\bfseries 07}
  (2012) 168},
\href{http://arxiv.org/abs/1204.0519}{{\ttfamily arXiv:1204.0519 [hep-th]}}.

\bibitem{Donos:2013eha}
A.~Donos and J.~P. Gauntlett, ``{Holographic Q-lattices},''
  \href{http://dx.doi.org/10.1007/JHEP04(2014)040}{{\em JHEP} {\bfseries 04}
  (2014) 040},
\href{http://arxiv.org/abs/1311.3292}{{\ttfamily arXiv:1311.3292 [hep-th]}}.

\bibitem{Andrade:2013gsa}
T.~Andrade and B.~Withers, ``{A simple holographic model of momentum
  relaxation},'' \href{http://dx.doi.org/10.1007/JHEP05(2014)101}{{\em JHEP}
  {\bfseries 05} (2014) 101},
\href{http://arxiv.org/abs/1311.5157}{{\ttfamily arXiv:1311.5157 [hep-th]}}.

\bibitem{Donos:2014yya}
A.~Donos and J.~P. Gauntlett, ``{The thermoelectric properties of inhomogeneous
  holographic lattices},''
  \href{http://dx.doi.org/10.1007/JHEP01(2015)035}{{\em JHEP} {\bfseries 1501}
  (2015) 035},
\href{http://arxiv.org/abs/1409.6875}{{\ttfamily arXiv:1409.6875 [hep-th]}}.

\bibitem{Donos:2021pkk}
A.~Donos, P.~Kailidis, and C.~Pantelidou, ``{Dissipation in holographic
  superfluids},'' \href{http://dx.doi.org/10.1007/JHEP09(2021)134}{{\em JHEP}
  {\bfseries 09} (2021) 134}, \href{http://arxiv.org/abs/2107.03680}{{\ttfamily
  arXiv:2107.03680 [hep-th]}}.

\bibitem{Ammon:2021pyz}
M.~Ammon, D.~Arean, M.~Baggioli, S.~Gray, and S.~Grieninger,
  ``{Pseudo-spontaneous $U(1)$ symmetry breaking in hydrodynamics and
  holography},'' \href{http://dx.doi.org/10.1007/JHEP03(2022)015}{{\em JHEP}
  {\bfseries 03} (2022) 015}, \href{http://arxiv.org/abs/2111.10305}{{\ttfamily
  arXiv:2111.10305 [hep-th]}}.

\bibitem{Heller:2013fn}
M.~P. Heller, R.~A. Janik, and P.~Witaszczyk, ``{Hydrodynamic Gradient
  Expansion in Gauge Theory Plasmas},''
  \href{http://dx.doi.org/10.1103/PhysRevLett.110.211602}{{\em Phys. Rev.
  Lett.} {\bfseries 110} no.~21, (2013) 211602},
  \href{http://arxiv.org/abs/1302.0697}{{\ttfamily arXiv:1302.0697 [hep-th]}}.

\bibitem{Heller:2015dha}
M.~P. Heller and M.~Spalinski, ``{Hydrodynamics Beyond the Gradient Expansion:
  Resurgence and Resummation},''
  \href{http://dx.doi.org/10.1103/PhysRevLett.115.072501}{{\em Phys. Rev.
  Lett.} {\bfseries 115} no.~7, (2015) 072501},
  \href{http://arxiv.org/abs/1503.07514}{{\ttfamily arXiv:1503.07514
  [hep-th]}}.

\bibitem{Withers:2018srf}
B.~Withers, ``{Short-lived modes from hydrodynamic dispersion relations},''
  \href{http://dx.doi.org/10.1007/JHEP06(2018)059}{{\em JHEP} {\bfseries 06}
  (2018) 059}, \href{http://arxiv.org/abs/1803.08058}{{\ttfamily
  arXiv:1803.08058 [hep-th]}}.

\bibitem{Grozdanov:2019kge}
S.~Grozdanov, P.~K. Kovtun, A.~O. Starinets, and P.~Tadi{\'c}, ``{Convergence
  of the Gradient Expansion in Hydrodynamics},''
  \href{http://dx.doi.org/10.1103/PhysRevLett.122.251601}{{\em Phys. Rev.
  Lett.} {\bfseries 122} no.~25, (2019) 251601},
\href{http://arxiv.org/abs/1904.01018}{{\ttfamily arXiv:1904.01018 [hep-th]}}.

\bibitem{Grozdanov:2019uhi}
S.~Grozdanov, P.~K. Kovtun, A.~O. Starinets, and P.~Tadi\'c, ``{The complex
  life of hydrodynamic modes},''
  \href{http://dx.doi.org/10.1007/JHEP11(2019)097}{{\em JHEP} {\bfseries 11}
  (2019) 097}, \href{http://arxiv.org/abs/1904.12862}{{\ttfamily
  arXiv:1904.12862 [hep-th]}}.

\bibitem{Son:2007vk}
D.~T. Son and A.~O. Starinets, ``{Viscosity, Black Holes, and Quantum Field
  Theory},''
  \href{http://dx.doi.org/10.1146/annurev.nucl.57.090506.123120}{{\em Ann. Rev.
  Nucl. Part. Sci.} {\bfseries 57} (2007) 95--118},
\href{http://arxiv.org/abs/0704.0240}{{\ttfamily arXiv:0704.0240 [hep-th]}}.

\bibitem{Bhattacharyya:2008jc}
S.~Bhattacharyya, V.~E. Hubeny, S.~Minwalla, and M.~Rangamani, ``{Nonlinear
  Fluid Dynamics from Gravity},''
  \href{http://dx.doi.org/10.1088/1126-6708/2008/02/045}{{\em JHEP} {\bfseries
  0802} (2008) 045},
\href{http://arxiv.org/abs/0712.2456}{{\ttfamily arXiv:0712.2456 [hep-th]}}.

\bibitem{Haack:2008cp}
M.~Haack and A.~Yarom, ``{Nonlinear viscous hydrodynamics in various dimensions
  using AdS/CFT},'' \href{http://dx.doi.org/10.1088/1126-6708/2008/10/063}{{\em
  JHEP} {\bfseries 10} (2008) 063},
\href{http://arxiv.org/abs/0806.4602}{{\ttfamily arXiv:0806.4602 [hep-th]}}.

\bibitem{Erdmenger:2008rm}
J.~Erdmenger, M.~Haack, M.~Kaminski, and A.~Yarom, ``{Fluid dynamics of
  R-charged black holes},''
  \href{http://dx.doi.org/10.1088/1126-6708/2009/01/055}{{\em JHEP} {\bfseries
  01} (2009) 055},
\href{http://arxiv.org/abs/0809.2488}{{\ttfamily arXiv:0809.2488 [hep-th]}}.

\bibitem{Banerjee:2008th}
N.~Banerjee, J.~Bhattacharya, S.~Bhattacharyya, S.~Dutta, R.~Loganayagam, and
  P.~Surowka, ``{Hydrodynamics from charged black branes},''
  \href{http://dx.doi.org/10.1007/JHEP01(2011)094}{{\em JHEP} {\bfseries 01}
  (2011) 094}, \href{http://arxiv.org/abs/0809.2596}{{\ttfamily arXiv:0809.2596
  [hep-th]}}.

\bibitem{Iqbal:2008by}
N.~Iqbal and H.~Liu, ``{Universality of the hydrodynamic limit in AdS/CFT and
  the membrane paradigm},''
  \href{http://dx.doi.org/10.1103/PhysRevD.79.025023}{{\em Phys.Rev.}
  {\bfseries D79} (2009) 025023},
\href{http://arxiv.org/abs/0809.3808}{{\ttfamily arXiv:0809.3808 [hep-th]}}.

\bibitem{Donos:2014cya}
A.~Donos and J.~P. Gauntlett, ``{Thermoelectric DC conductivities from black
  hole horizons},'' \href{http://dx.doi.org/10.1007/JHEP11(2014)081}{{\em JHEP}
  {\bfseries 1411} (2014) 081},
\href{http://arxiv.org/abs/1406.4742}{{\ttfamily arXiv:1406.4742 [hep-th]}}.

\bibitem{Donos:2015gia}
A.~Donos and J.~P. Gauntlett, ``{Navier-Stokes Equations on Black Hole Horizons
  and DC Thermoelectric Conductivity},''
  \href{http://dx.doi.org/10.1103/PhysRevD.92.121901}{{\em Phys. Rev.}
  {\bfseries D92} no.~12, (2015) 121901},
\href{http://arxiv.org/abs/1506.01360}{{\ttfamily arXiv:1506.01360 [hep-th]}}.

\bibitem{Donos:2015bxe}
A.~Donos, J.~P. Gauntlett, T.~Griffin, and L.~Melgar, ``{DC Conductivity of
  Magnetised Holographic Matter},''
  \href{http://dx.doi.org/10.1007/JHEP01(2016)113}{{\em JHEP} {\bfseries 01}
  (2016) 113},
\href{http://arxiv.org/abs/1511.00713}{{\ttfamily arXiv:1511.00713 [hep-th]}}.

\bibitem{Donos:2017oym}
A.~Donos, J.~P. Gauntlett, T.~Griffin, and L.~Melgar, ``{DC Conductivity and
  Higher Derivative Gravity},''
  \href{http://dx.doi.org/10.1088/1361-6382/aa744a}{{\em Class. Quant. Grav.}
  {\bfseries 34} no.~13, (2017) 135015},
\href{http://arxiv.org/abs/1701.01389}{{\ttfamily arXiv:1701.01389 [hep-th]}}.

\bibitem{Eling:2011ms}
C.~Eling and Y.~Oz, ``{A Novel Formula for Bulk Viscosity from the Null Horizon
  Focusing Equation},'' \href{http://dx.doi.org/10.1007/JHEP06(2011)007}{{\em
  JHEP} {\bfseries 06} (2011) 007},
  \href{http://arxiv.org/abs/1103.1657}{{\ttfamily arXiv:1103.1657 [hep-th]}}.

\bibitem{Eling:2009pb}
C.~Eling, I.~Fouxon, and Y.~Oz, ``{The Incompressible Navier-Stokes Equations
  From Membrane Dynamics},''
  \href{http://dx.doi.org/10.1016/j.physletb.2009.09.028}{{\em Phys. Lett.}
  {\bfseries B680} (2009) 496--499},
\href{http://arxiv.org/abs/0905.3638}{{\ttfamily arXiv:0905.3638 [hep-th]}}.

\bibitem{Crnkovic:1986ex}
C.~Crnkovic and E.~Witten, ``{Covariant description of canonical formalism in
  geometrical theories},''.

\bibitem{Grant:2005qc}
L.~Grant, L.~Maoz, J.~Marsano, K.~Papadodimas, and V.~S. Rychkov,
  ``{Minisuperspace quantization of 'Bubbling AdS' and free fermion
  droplets},'' \href{http://dx.doi.org/10.1088/1126-6708/2005/08/025}{{\em
  JHEP} {\bfseries 08} (2005) 025},
  \href{http://arxiv.org/abs/hep-th/0505079}{{\ttfamily arXiv:hep-th/0505079}}.

\bibitem{Maoz:2005nk}
L.~Maoz and V.~S. Rychkov, ``{Geometry quantization from supergravity: The Case
  of 'Bubbling AdS'},''
  \href{http://dx.doi.org/10.1088/1126-6708/2005/08/096}{{\em JHEP} {\bfseries
  08} (2005) 096}, \href{http://arxiv.org/abs/hep-th/0508059}{{\ttfamily
  arXiv:hep-th/0508059}}.

\bibitem{Donos:2005vs}
A.~Donos and A.~Jevicki, ``{Dynamics of chiral primaries in AdS(3) x S**3 x
  T**4},'' \href{http://dx.doi.org/10.1103/PhysRevD.73.085010}{{\em Phys. Rev.
  D} {\bfseries 73} (2006) 085010},
  \href{http://arxiv.org/abs/hep-th/0512017}{{\ttfamily arXiv:hep-th/0512017}}.

\bibitem{Wald:1993nt}
R.~M. Wald, ``{Black hole entropy is the Noether charge},''
  \href{http://dx.doi.org/10.1103/PhysRevD.48.R3427}{{\em Phys. Rev.}
  {\bfseries D48} no.~8, (1993) R3427--R3431},
\href{http://arxiv.org/abs/gr-qc/9307038}{{\ttfamily arXiv:gr-qc/9307038
  [gr-qc]}}.

\bibitem{Iyer:1994ys}
V.~Iyer and R.~M. Wald, ``{Some properties of Noether charge and a proposal for
  dynamical black hole entropy},''
  \href{http://dx.doi.org/10.1103/PhysRevD.50.846}{{\em Phys. Rev.} {\bfseries
  D50} (1994) 846--864},
\href{http://arxiv.org/abs/gr-qc/9403028}{{\ttfamily arXiv:gr-qc/9403028
  [gr-qc]}}.

\bibitem{Papadimitriou:2005ii}
I.~Papadimitriou and K.~Skenderis, ``{Thermodynamics of asymptotically locally
  AdS spacetimes},''
  \href{http://dx.doi.org/10.1088/1126-6708/2005/08/004}{{\em JHEP} {\bfseries
  0508} (2005) 004},
\href{http://arxiv.org/abs/hep-th/0505190}{{\ttfamily arXiv:hep-th/0505190
  [hep-th]}}.

\bibitem{Skenderis:2002wp}
K.~Skenderis, ``{Lecture notes on holographic renormalization},''
  \href{http://dx.doi.org/10.1088/0264-9381/19/22/306}{{\em Class.Quant.Grav.}
  {\bfseries 19} (2002) 5849--5876},
\href{http://arxiv.org/abs/hep-th/0209067}{{\ttfamily arXiv:hep-th/0209067
  [hep-th]}}.

\bibitem{Donos:2013cka}
A.~Donos and J.~P. Gauntlett, ``{On the thermodynamics of periodic AdS black
  branes},'' \href{http://dx.doi.org/10.1007/JHEP10(2013)038}{{\em JHEP}
  {\bfseries 1310} (2013) 038},
\href{http://arxiv.org/abs/1306.4937}{{\ttfamily arXiv:1306.4937 [hep-th]}}.

\bibitem{Donos:2022xfd}
A.~Donos and C.~Pantelidou, ``{Higgs/amplitude mode dynamics from
  holography},'' \href{http://dx.doi.org/10.1007/JHEP08(2022)246}{{\em JHEP}
  {\bfseries 08} (2022) 246}, \href{http://arxiv.org/abs/2205.06294}{{\ttfamily
  arXiv:2205.06294 [hep-th]}}.

\bibitem{Herzog:2011ec}
C.~P. Herzog, N.~Lisker, P.~Surowka, and A.~Yarom, ``{Transport in holographic
  superfluids},'' \href{http://dx.doi.org/10.1007/JHEP08(2011)052}{{\em JHEP}
  {\bfseries 08} (2011) 052},
\href{http://arxiv.org/abs/1101.3330}{{\ttfamily arXiv:1101.3330 [hep-th]}}.

\bibitem{Bhattacharya:2011eea}
J.~Bhattacharya, S.~Bhattacharyya, and S.~Minwalla, ``{Dissipative Superfluid
  dynamics from gravity},''
  \href{http://dx.doi.org/10.1007/JHEP04(2011)125}{{\em JHEP} {\bfseries 1104}
  (2011) 125},
\href{http://arxiv.org/abs/1101.3332}{{\ttfamily arXiv:1101.3332 [hep-th]}}.

\bibitem{Buchel:2009mf}
A.~Buchel and C.~Pagnutti, ``{Transport at criticality},''
  \href{http://dx.doi.org/10.1016/j.nuclphysb.2010.03.016}{{\em Nucl. Phys.}
  {\bfseries B834} (2010) 222--236},
\href{http://arxiv.org/abs/0912.3212}{{\ttfamily arXiv:0912.3212 [hep-th]}}.

\end{thebibliography}\endgroup
\end{document}